\documentclass[10pt]{iopart}
\usepackage{iopams}
\expandafter\let\csname equation*\endcsname\relax
\expandafter\let\csname endequation*\endcsname\relax
\usepackage{amsmath}
\usepackage{siunitx}
\usepackage{braket}

\usepackage{bm,graphicx,subfigure,epsfig}
\graphicspath{{Pictures/}}
\usepackage{color}
\usepackage[T1]{fontenc}
\usepackage{cuted}

\usepackage{marginnote}
\usepackage{array,multirow}
\usepackage{mathtools}
\usepackage{tabularx,booktabs}

\newcolumntype{P}[1]{>{\centering\arraybackslash}p{#1}}

\newcolumntype{Y}{>{\centering\arraybackslash}X}
\newcommand{\vk}{\ensuremath{\mathbf{k}}}
\newcommand{\vp}{\ensuremath{\mathbf{p}}}
\newcommand{\vq}{\ensuremath{\mathbf{q}}}
\newcommand{\dr}{\ensuremath{\mathbf{r}}}
\newcommand\thickvrule[1][2pt]{\vrule width #1}

\begin{document}

\title{Interacting Dirac Materials}

\author{S. Banerjee$^{1,2,3}$, D.S.L. Abergel$^2$, H. {\AA}gren$^{3,4}$, G. Aeppli$^{5,6,7}$, A.V. Balatsky$^{2,8}$}

\vspace{2.5mm}
\address{$^1$ Theoretical Physics III, Center for Electronic Correlations and Magnetism, Institute of Physics, University of Augsburg, 86135 Augsburg Germany}
\address{$^2$Nordita, KTH Royal Institute of Technology and Stockholm University, Roslagstullsbacken 23, 10691 Stockholm, Sweden}
\address{$^3$Division of Theoretical Chemistry and Biology, Royal Institute of Technology, SE-10691 Stockholm, Sweden}
\address{$^4$ Department of Physics and Astronomy, Uppsala University, Box 516, SE-751 20 Uppsala, Sweden}
\address{$^5$ Paul Scherrer Institute, CH-5232 Villigen PSI, Switzerland}
\address{$^6$ Laboratory for Solid State Physics, ETH Zurich, Zurich, CH-8093, Switzerland}
\address{$^7$ Institut de Physique, EPF Lausanne, Lausanne, CH-1015, Switzerland}
\address{$^8$ Department of Physics, University of Connecticut, Storrs, Connecticut 06269, USA}

\ead{\mailto{saikat.banerjee@physik.uni-augsburg.de}}
\vspace{10pt}

\begin{indented}
\item[]\today
\end{indented}

\begin{abstract}
We investigate the extent to which the class of Dirac materials in two-dimensions provides general statements about the behavior of both fermionic and bosonic Dirac quasiparticles in the interacting regime. For both quasiparticle types, we find common features for the interaction induced renormalization of the conical Dirac spectrum. We perform the perturbative renormalization analysis and compute the self-energy for both quasiparticle types with different interactions and collate previous results from the literature whenever necessary. Guided by the systematic presentation of our results in Table~\ref{Summary}, we conclude that long-range interactions generically lead to an increase of the slope of the single-particle Dirac cone, whereas short-range interactions lead to a decrease. The quasiparticle statistics does not qualitatively impact the self-energy correction for long-range repulsion but does affect the behavior of short-range coupled systems, giving rise to different thermal power-law contributions. The possibility of a universal description of the Dirac materials based on these features is also mentioned.
\end{abstract}

\vspace{2pc}

\submitto{\JPCM}
\maketitle
\ioptwocol

\section{Introduction}

The discovery of graphene~\cite{Novoselov2005,Geim2007,RevModPhys.81.109} and other materials with Dirac nodes has led to a strong interest in materials with Dirac dispersion~\cite{Balatsky,Vafek,Jinying} in condensed matter physics. In a typical $d$-dimensional metal, the Fermi surface is a $(d-1)$-dimensional manifold. In Dirac materials, on the other hand, the locus of points with zero quasiparticle energy shrinks to  $d-2$ or $d-3$ dimensions due to some additional symmetries present in the system~\cite{Balatsky, Jinying}. Specifically, for two-dimensional (2D) Dirac materials, Dirac nodes have to be point objects. The low-energy quasiparticles around this Fermi point have linear dispersion and are described by a massless Dirac-like equation. Because of the linear dispersion, the quasiparticles behave like relativistic particles with an effective speed of light replaced by the Fermi velocity. Many realizations of Dirac fermions have been discussed in, for example, superfluid phases of $^3$He~\cite{doi:10.1142/1439}, the surface states of topological insulators~\cite{RevModPhys.82.3045, RevModPhys.83.1057, ChenY}, high-temperature $d$-wave superconductors~\cite{PhysRevLett.113.046401}, graphene~\cite{Abergel} and several artificial electronic systems~\cite{Gomes2012, Polini2013}. The resulting Dirac cone in the spectrum is protected by symmetries~\cite{PhysRevLett.115.126803} and is robust under disorder and interaction. For example, Dirac nodes are protected by parity and time-reversal symmetries in graphene~\cite{RevModPhys.81.109}, by time-reversal and mirror symmetries in topological insulators~\cite{RevModPhys.82.3045} and by a combination of the $C_4$  rotation and gauge transformation in the case of d-wave superconductors in two-dimensional square lattices~\cite{Balatsky}. The Dirac nodes lead to commonly observed universal features of excitations such as the scaling of the Landau levels with the magnetic field as $\sqrt{B}$~\cite{RevModPhys.81.109}, universal metallic and thermal conductivities related to the nodal structure of carriers, suppressed back-scattering~\cite{Balatsky}, and Klein tunneling~\cite{Katsnelson2006}. The common properties shared by these systems bring about the unified concept of Dirac materials as a useful category for explaining their behavior.

\begin{figure}[t]
\centering
\includegraphics[width=1\linewidth]{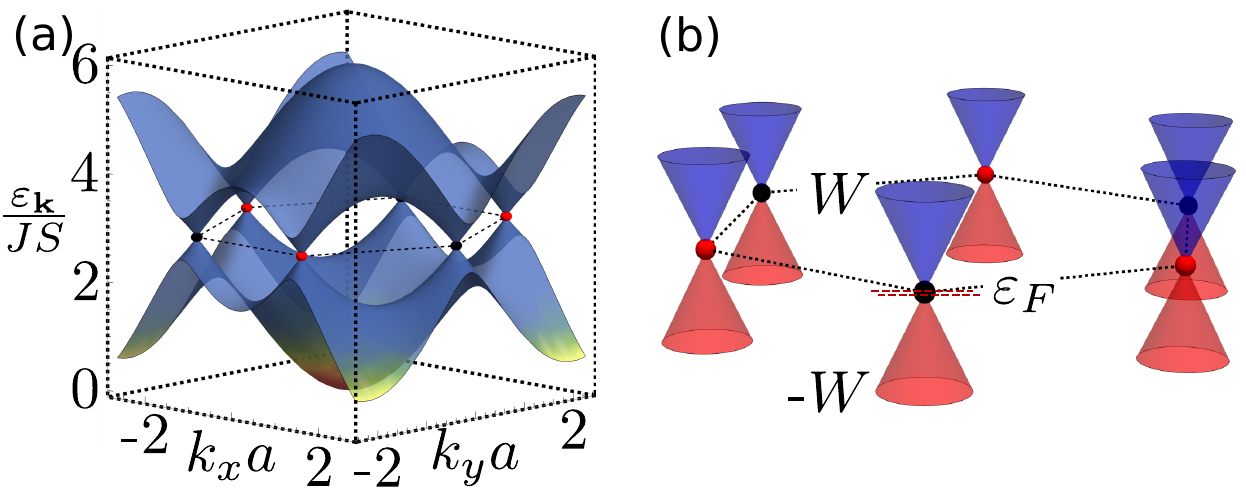}
\caption{Comparison of the band-structure for non-interacting Dirac bosons (left panel) vs fermions (right panel). (a) The energy spectrum for a single magnon excitation induced by a ferromagnetic Heisenberg model defined on a honeycomb lattice (see Eq.~\ref{eq.23} and Refs.~\cite{Jonas,Saikat}). The magnons can occupy any state belonging to the two bands which cross each other at the Dirac points (marked in red and black points). The near vicinity area of the spectrum around these crossing points can be effectively described by a Dirac-like Hamiltonian. (b) In contrast, the Dirac fermionic excitations are described by the low-lying states from the Fermi energy $\varepsilon_F$ at the Dirac node within a bandwidth $\pm W$.}
\label{fig:Spectrum}
\end{figure}

Recently, we are witnessing an upsurge of theoretical and experimental investigations in bosonic systems possessing Dirac-like spectra. Plausible physical realizations of such bosonic Dirac materials range from artificial honeycomb lattices made out of superconducting grains~\cite{Banerjee} to magnets~\cite{Jonas}, photonic and acoustic crystals~\cite{Raghu,Dubois}, and plasmonic devices~\cite{Basov}. At the level of the single-particle physics, the band structures of bosonic and fermionic Dirac materials bear interesting \textit{similarities}: both have symmetry protected band crossing points around which the excitations have linear dispersion parameterized by a velocity. However, for \textit{fermions}, these are typically low-energy \textit{excitations} around the Fermi energy at the \textit{Dirac} nodes; whereas the \textit{bosonic Dirac excitations} lie at a higher energy from the ground state (see Fig.~\ref{fig:Spectrum}).

Several important questions about this new class of materials remain unanswered: is there a universal behavior for bosonic and fermionic Dirac excitations when interactions are taken into account? What role does the form of the interaction play? And how are these results modified at finite temperature? The first question is particularly important because the quasiparticle statistics will have a strong impact on interacting behavior. Our results in Sec.~\ref{sec:interactions}, taken together with the existing literature in the field of Dirac materials, allows us to extract certain striking similarities between the renormalization of fermionic and bosonic Dirac materials. On a qualitative level, long-range interactions (Coulomb type) lead to an increase in the slope of the Dirac cone, whereas short-range interactions decrease this slope. Furthermore, both the fermionic and bosonic excitations acquire logarithmic temperature-dependent self-energy corrections (see  Table~\ref{Summary}). Such temperature dependence arises from Dirac dispersions in the fermionic illustration, whereas for bosons it is associated with the typical parabolic dispersions. On the other hand, different power-law corrections in temperature and momentum are obtained for the renormalized dispersions for short-range interactions.

Table ~\ref{Summary} summarizes the trends in renormalization of Dirac node velocity. These trends  indicate  common features in  velocity dependence in short and long range interacting cases. One indeed might suspect there are universal trends in renormalzations of Dirac nodes depending on dimensionality and nature of  interaction. To determine if indeed these trends are universal one would need to investigate a broader class of models in D = 1,2,3. This investigation of broader class of models is left for the future work.

%=================================================================================================================%=================================================================================================================
%=================================================================================================================%=================================================================================================================

In this paper, we analyze the renormalized dispersions and the induced quasiparticle lifetimes for the weakly interacting two-dimenisonal Dirac excitations. We therefore set up our models for : (a) \textit{Dirac fermions} interacting via an \textit{onsite Hubbard} repulsion and (b) \textit{Dirac bosons} subject to a \textit{long-range} Coulomb interaction. We purposefully recollect the relevant results for the interacting Dirac magnons (see Ref.~\cite{Saikat}) as a prototype for Dirac bosons coupled via short-range interaction. In a similar spirit, any explicit analysis for the renormalization of the Dirac fermions subject to long-range interactions is dropped. Instead, we rely on the well-known results obtained in Refs.~\cite{RevModPhys.81.109,GONZALEZ1994595}. We do not consider the role of spin-orbit coupling or similar interactions that might lead to a gap opening at the Dirac point, and consequently, trigger a topological phase transition. In this paper, we focus in the systems where the Dirac crossing point is preserved. Only, the linear spectrum around the crossing point is modified due to various interactions. Hence, interaction induced renormalization of Dirac spectrum does not change the topology of the underlying system.

The rest of the paper is structured as follows. We briefly revisit the free Dirac theory of bosons and fermions in Sec.~\ref{sec:singleparticle}. Sec.~\ref{sec:interactions} is divided into multiple subsections, where we systematically explore the individual scenarios as discussed in the previous paragraph. Although, the main focus of this paper is on two-dimensional materials, we provide a short discussion on the interacting Dirac bosons in one and three dimensions in Sec.~\ref{sec:db_dimension}. We finally provide a detailed discussion for our results on the interacting Dirac matter and conclude in Sec.~\ref{sec:discussion}.

\section{Dirac Materials: Noninteracting quasiparticles \label{sec:singleparticle}}

 We begin by revisiting the Dirac theory of non-interacting quasiparticles. Numerous material-specific realizations of the Dirac excitations of both fermionic and bosonic origin have been observed in the last two decades. Ref.~\cite{Balatsky} provides an interesting review of various fermionic Dirac excitations; whereas various Refs.~\cite{Banerjee,Jonas,Raghu,Dubois,Basov} (and more) investigated the bosonic counterparts. Band-structures like the ones shown in Fig.~\ref{fig:Spectrum} containing Dirac-like nodes typically arise from a tight-binding model defined on the non-symmorphic lattice such as honeycomb, kagome, etc. A generic Hamiltonian, in such a case, can be written as
\begin{equation}\label{eq.1}
H=  -t \sum_{\langle ij \rangle} c^{\dagger}_{i}c_{j}(b^{\dagger}_{i}b_{j}) + h.c.  \mp \varepsilon_F(\varepsilon_B) \sum_{i} c^{\dagger}_ic_i ( b^{\dagger}_ib_i),
\end{equation}
where $c^{\dagger}_i (b^{\dagger}_i)$ is the creation operator for the fermionic (bosonic) quasiparticles, like electrons (magnons, phonons, plasmons, etc.), at site $i$ within the unit cell of the lattice. $t$- is the amplitude for the quasiparticle hopping between the nearest-neighbor sites $\langle ij \rangle$. The lattice constant is denoted by $a$. $\varepsilon_F (\varepsilon_B)$ denotes the Fermi (an onsite) energy for fermionic (bosonic) quasiparticles.

For example, in a honeycomb lattice, the Hamiltonian in Eq.~\ref{eq.1} gives rise to band-structures as shown in Fig.~\ref{fig:Spectrum}. In particular, for magnons (\textit{bosonic excitations}, $t = JS$), one obtains a band-structure as in Fig.~\ref{fig:Spectrum}a (see Ref.~\cite{Saikat} for the details). The onsite energy $\varepsilon_B (= 3JS)$ in this case, accounts for the positivity of the magnonic dispersions. In contrast, for \textit{fermionic excitations}, $\varepsilon_F$ is typically the chemical potential at the nodal point. The low-lying fermionic states, within the bandwidth $\pm W$ around the nodal points (the red/black points in Fig.~\ref{fig:Spectrum}b), are typically described by an effective Dirac Hamiltonian
\begin{equation}\label{eq.2}
H_{eff} \simeq \sum_{\vk} {\bm\Psi}^{\dagger}_{\vk} {\bm H}_\vk {\bm\Psi}_{\vk}; \quad {\bm H}_\vk = v_D {\bm \sigma} \cdot \vk,
\end{equation}
where $v_D$, the slope of the cone, is the Dirac velocity. The spinor ${\bm\Psi}^{\dagger}_{\vk}$ is composed of the states belonging to the two inequivalent sub-lattices in the honeycomb unit cell. The Pauli matrices $\sigma_{x,y}$ correspond to the \textit{pseudospin} structure arising out of the sub-lattice degrees of freedom. See~\ref{sec:small_deri} for the details of the derivation of fermionic and bosonic Dirac Hamiltonian.

%=================================================================================================================================
\begin{table*}[tb]
\caption{This table summarizes the primary results for various self-energy corrections in the case of interacting Dirac materials with both type of quasiparticles: fermions and bosons. The self-energy $\Sigma$ corresponds to the upward region of the band-structure around the Dirac cone (see Fig.~\ref{fig:Spectrum}). Here, $x$ corresponds to either a single or a combination of the variables \--- momentum $k$, frequency $\omega$ or temperature $T$ (in the unit of Boltzmann constant, $k_B = 1$; $\hbar = 1$).  Note: The respective referrences for the known results are provided in the entries where-ever applicable. The pseudo-relativistic invariance of the Dirac fermions has been used to deduce the momentum dependence of $\Sigma$ in case of the Hubbard repulsion. At zero temperature, the bosonic quasiparticle excitations are absent and hence do not lead to any interactions. Hence, the self-energy correction is absent in these cases. However, this will not hold for a generic bosonic system.}\label{Summary}
\bigskip
\centering
\setlength{\tabcolsep}{0.7em}
\renewcommand{\arraystretch}{2}
\begin{tabular}{|c|c|c|c|c|}
\cline{1-5} \hline \hline
\multicolumn{1}{|c|}{Quantity}
 &\multicolumn{2}{|c|}{Quasiparticle} & \multicolumn{1}{c!{\thickvrule}}{Long-range interaction} & \multicolumn{1}{c|}{Short-range interaction} \\  \cline{1-5} \hline \hline
\multirow{4}{*}{Re$\Sigma(x)$} & \multirow{2}{*}{Fermion} & {$T=0$} &\multicolumn{1}{c!{\thickvrule}}{$k\log\frac{\Lambda}{k}$, $\omega \log \omega$~\cite{PhysRevB.75.121406}}  & {$\sim$-$U^2$max($k^3,\omega^3$)}\\ \cline{3-5}
&{} & {$T\neq0$}  & \multicolumn{1}{c!{\thickvrule}}{$k\log\Big[\frac{\Lambda}{\text{max}(k,T)}\Big]$}  &  {-$\text{max}(k^3, \omega^3)$}\\ \cline{2-5}
& \multirow{2}{*}{Boson} & {$T=0$}  & \multicolumn{1}{c!{\thickvrule}}{Zero} &  {Zero~\cite{Saikat}}\\ \cline{3-5}
&{}  & {$T\neq0$}  & \multicolumn{1}{c!{\thickvrule}}{$T\log T$} &  {-$T^2$ (see Fig.~\ref{fig:renorm_spectrum})~\cite{Saikat}}\\ \hline \hline
\multirow{4}{*}{Im$\Sigma(x)$} & \multirow{2}{*}{Fermion} & {$T=0$} &\multicolumn{1}{c!{\thickvrule}}{-max($|k|$,$|\omega|$)~\cite{PhysRevB.75.121406}}  & {-max($k^4,\omega^4$)}\\ \cline{3-5}
&{} & {$T\neq0$}  & \multicolumn{1}{c!{\thickvrule}}{-$\text{max}(|k|,T,|\omega|)$}  & {-$\text{max}(k^4, \omega^4,T^4)$} \\ \cline{2-5}
& \multirow{2}{*}{Boson} & {$T=0$}  & \multicolumn{1}{c!{\thickvrule}}{Zero} & {Zero~\cite{Saikat}} \\ \cline{3-5}
&{}  & {$T\neq0$}  & \multicolumn{1}{c!{\thickvrule}}{-$T\log T$}  & {-$T^2$ (see Fig.~\ref{fig:Fig3})~\cite{Saikat}}\\ \hline \hline
\end{tabular}
\end{table*}
%=================================================================================================================================
We notice that the bosonic Dirac excitations are high-energy states compared to the band-bottom edge in Fig.~\ref{fig:Spectrum}a. As a result, it is sufficient to focus on the conical portion (see Fig.~\ref{fig:Spectrum}b) around the Dirac nodes for fermions, whereas for bosons the entire band-structure is considered as illustrated in Fig.~\ref{fig:Spectrum}a. Despite these mutual differences, Dirac bosons and fermions display various common experimental features \--- for example, the Klein-tunneling in photons possessing Dirac-like dispersions of topological photonic insulators~\cite{Nieaap8802}, the square-root magnetic-field-dependent Landau quantization in Dirac phonons~\cite{Wen2019}. Motivated by these common experimental observations, we set out to investigate whether such similarities exist in the interacting regime.

\section{Interacting Dirac materials \label{sec:interactions}}

So far, we focused on the non-interacting Dirac materials. Despite, their mutual differences for the quasiparticle type, both fermionic and bosonic Dirac materials display numerous identical experimental signatures. However, whether these similarities indicate an underlying universal description, especially in the interacting regime, remains unexplored. In the subsequent sections, we focus on a subset of interacting Dirac systems and try to extract some common features.
%=================================================================================================================================
\begin{figure*}[t]
\centering
\includegraphics[width=1\linewidth]{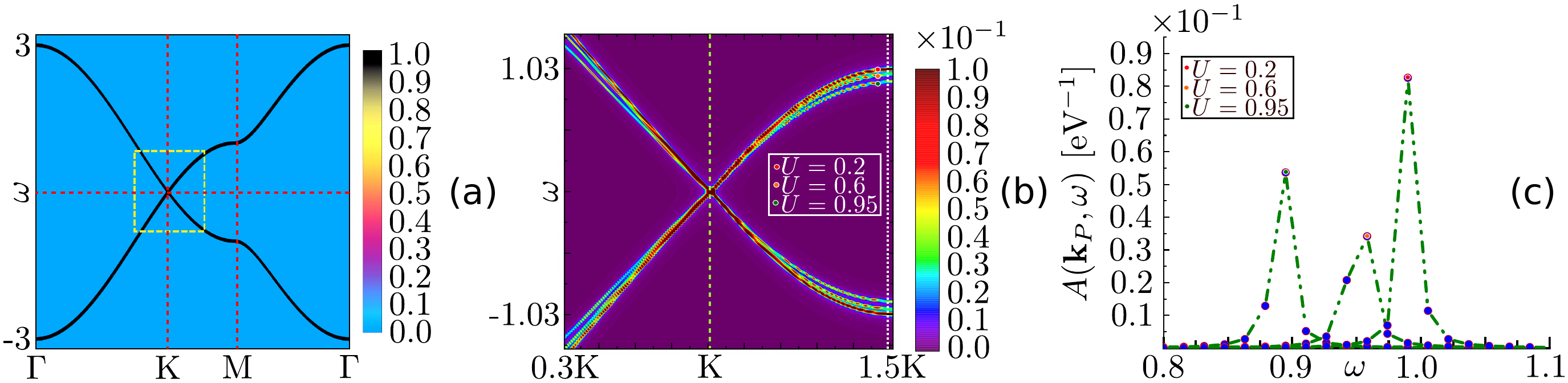}
\caption{(a) The spectral function $A(\vk ,\omega)$ (in units of eV$^{-1}$) for the non-interacting fermions on a honeycomb lattice is shown along the high-symmetry points in the Brillouin zone. The linear Dirac spectrum (with $U=0$ in Eq.~\ref{eq.10}) is highlighted by the boxed area. (b) The renormalized spectral function $A(\vk ,\omega)$ (in units eV$^{-1}$) at zero temperature for three different strengths of the Hubbard interaction $U = 0.2,\; 0.6,\; \&\; 0.95$ (marked by filled circles) as mentioned in the panel. (c) The spectral function, at fixed momentum $\mathbf{k}_P$ (marked by the dotted white line in panel (b)), as function of the energy (positive valued) for various $U$ values as shown in the panel. Note: the tight-binding hopping $t$ is assumed to be 1 eV.}
\label{fig:Fig4}
\end{figure*}
%=================================================================================================================================
\subsection{Coulomb repulsion: Fermions \label{coulomb_fermi}}

We begin with the most popularly studied system: Dirac fermions interacting via a Coulomb repulsion $V_\vq = e^2/\epsilon_0q$. We witnessed an abundance of both theoretical and experimental investigations in this case, over the past twenty years (see e.g. Refs.~\cite{RevModPhys.81.109,GONZALEZ1994595,PhysRevB.89.235431,Elias}). Therefore, without including any explicit derivations, we recollect the results in this case as (see Ref.~\cite{RevModPhys.81.109} and at the end of~\ref{sec:AppHFT} for a brief description of the Hartree-Fock calculation leading to the logarithmic divergence)
\begin{equation}\label{eq.3}
{\bm H}_\vk \simeq v_D\left(1  + \alpha \log\frac{\Lambda}{k} \right) {\bm \sigma} \cdot \vk,
\end{equation}
where the second term in the bracket is the first-order self-energy correction to the non-interacting Dirac Hamiltonian. $\alpha = e^2/\epsilon_0 v_D$ is the  dimensionless coupling constant and $\Lambda$ ($\sim 2\pi/a$) is an ultraviolet cutoff. We notice that the renormalized Dirac velocity diverges as $k \to 0$ implying an increase in the Dirac velocity.

This divergence of the renormalized Dirac velocity at $\vk =0$, however, disappears at a finite temperature $T$ as
\begin{equation}\label{eq.4}
\delta v_D(\vk) =  \frac{e^2}{2\epsilon_0}\log \left( \frac{v_D\Lambda}{2k_BT} \right),
\end{equation}
where $k_B$ is the Boltzmann constant. The details of this result are provided in~\ref{sec:AppHFT}. Such a logarithmic temperature dependence has been previously mentioned in Ref.~\cite{PhysRevLett.99.226803} based on the scaling argument. Here, we provide an alternative derivation based on the perturbation theory. It is natural to expect that at low temperature, the Dirac velocity should renormalize as
\begin{equation}\label{eq.6}
\delta v_D(\vk) \propto \log \left( \frac{v_D\Lambda}{max\{v_D k, 2k_BT\}} \right).
\end{equation}

\subsection{Screened Coulomb interaction: Fermions \label{sc_coulomb_fermi}}

In the previous section, we discussed the renormalization of Dirac fermions coupled via the unscreened Coulomb interaction. In real samples  the bare Coulomb repulsion is screened. Such screening of Coulomb interaction in two-dimension has been previously derived in Refs.~\cite{giuliani_vignale_2005,PhysRevB.74.201401,RevModPhys.84.1067} as
\begin{equation}\label{eq.7}
V_{SC}(\dr) = \frac{e^2}{\epsilon_0 r} \big\{ 1 - \frac{\pi q_{TF}r}{2}\big[{\bf{H}}_0(q_{TF}r)-Y_0(q_{TF}r)\big]\big\},
\end{equation}
where $q_{TF} = 4 e^2 |\mu|/\epsilon_0 \hbar^2 v_D^2$ is the Thomas-Fermi wave-vector, $\mu$ is the chemical potential, and ${\bf{H}}_0$ and $Y_0$ are the Struve and Neumann functions. Following a similar analysis (see~\ref{sec:AppSY}) leading to the renormalized Dirac structure in Eq.~\ref{eq.3}, in this case we obtain
\begin{equation}\label{eq.8}
{\bm H}^{SC}_\vk \simeq v_D\left(1  + \frac{e^2}{ \epsilon_0v_D} \log\frac{\Lambda}{k +q_{TF}} \right) {\bm \sigma} \cdot \vk.
\end{equation}
The Dirac velocity, of course, acquires a similar logarithmic correction; however, the $\vk=0$ divergence disappears even at zero temperature. At finite temperature, we expect a similar modification to the renormalized Dirac velocity as in Eq.~\ref{eq.6} with $\delta v_D(\vk) \propto \log [v_D \Lambda/max\{v_D(k+q_{TF}),2k_BT\}]$. It becomes evident that the divergence of the Dirac velocity as in Eq.~\ref{eq.3}, arises out of the long-range nature of the unscreened Coulomb interaction. When this true long-range character is modified to an intermediate/short-range type, the divergence disappears as in Eq.~\ref{eq.8}.

In order to better understand this feature, we analyze the renormalization of the non-interacting Dirac Hamiltonian subject to a Yukawa potential $V_{Y}(\dr) = \frac{e^2}{\epsilon_0 r}e^{-q_{TF} r}$. Indeed, performing a calculation similar to the one in~\ref{sec:AppSY}, we obtain the renormalized Dirac Hamiltonian as
\begin{equation}\label{eq.9}
{\bm H}^{Y}_\vk \simeq v_D\left(1  + \frac{e^2}{ \epsilon_0v_D} \log\frac{\Lambda}{\sqrt{k^2+q_{TF}^2}} \right) {\bm \sigma} \cdot \vk,
\end{equation}
where, once again we notice that the renormalized Dirac velocity at $\vk = 0$ does not diverge due to the presence of the cut-off  $q_{TF}$, which is related to the finite range of the Yukawa potential.

\subsection{Onsite Hubbard interaction: Fermions \label{sec:hubbard_int}}

Following the previous two sections, we observed that a moderate/long-range interaction qualitatively leads to an upward renormalization of the Dirac velocity. To understand whether this generic feature extends over the entire range of the underlying interactions, here we focus on the opposite limit \---  an onsite Hubbard interaction.

 We perform the relevant calculations on a honeycomb lattice. The Hamiltonian can be written as
\begin{equation}\label{eq.10}
\mathcal{H} = -t \sum_{\langle ij \rangle,\sigma}c^{\dagger}_{i\sigma}c_{j\sigma} + U \sum_{i}n_{i\uparrow}n_{i\downarrow} - \epsilon_F \sum_{i,\sigma}c^{\dagger}_{i\sigma}c_{i\sigma},
\end{equation}
where $\sigma$ is the spin, and $U$ denotes the strength of the onsite repulsion. Other parameters are defined in Eq.~\ref{eq.1}. We assume $t = 1$ eV for further analysis. As discussed in Sec.~\ref{sec:singleparticle}, the non-interacting quasiparticles are described by a Dirac-like Hamiltonian at half-filling. Hubbard model on honeycomb lattice has been extensively studied in the past. Especially, the semi-metal to anti-ferromagnetic insulator transition has been the primary focus of the state-of-the-art numerical calculations~\cite{PhysRevB.91.165108,PhysRevX.6.011029,Sorella2012}. While much of investigations~\cite{Tang570,PhysRevLett.97.146401,PhysRevB.89.205128} focus on understanding such phase-transition, the semi-metallic phase for $U \ll U_c$ has remained relatively less explored. Here, $U_c$ represents the critical strength of the Hubbard interaction.

%===========================================================================
\begin{figure}[t]
\centering
\includegraphics[width=1\linewidth]{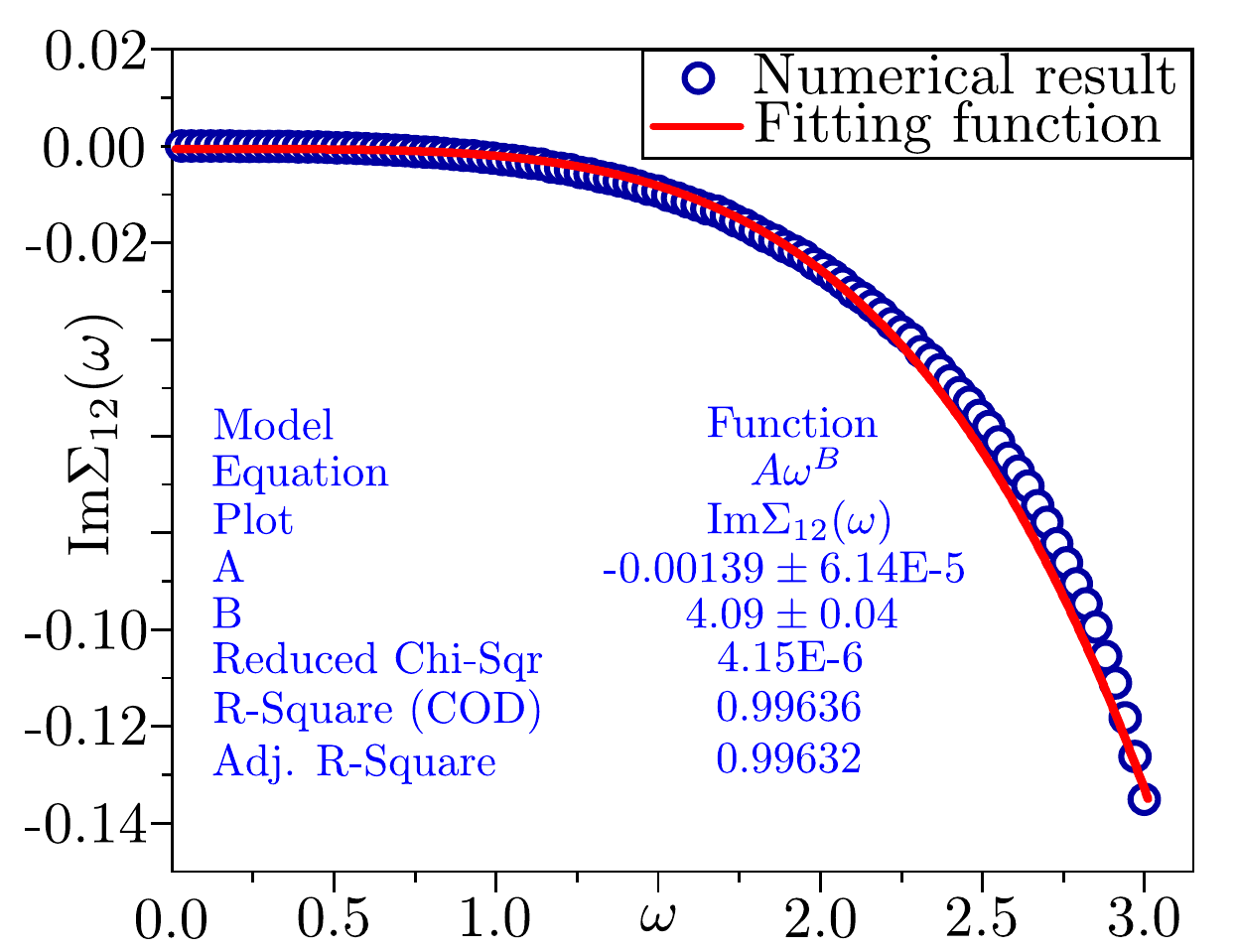}
\caption{The imaginary part (\textit{open circles}) of the off-diagonal self-energy $\Sigma_{12}$ (in the units of $t$) for the positive part of the frequency $\omega$ evaluated at the Dirac point at zero temperature for $U = 0.6t$. The numerical fit (\textit{solid red line}) of our result with the ansatz function $\text{Im}\Sigma_{12}(\omega) \sim A\cdot \omega^B$.}
\label{fig:MFL_Fit}
\end{figure}
%===========================================================================
%===========================================================================
\begin{figure}[b]
\centering
\includegraphics[width=1\linewidth]{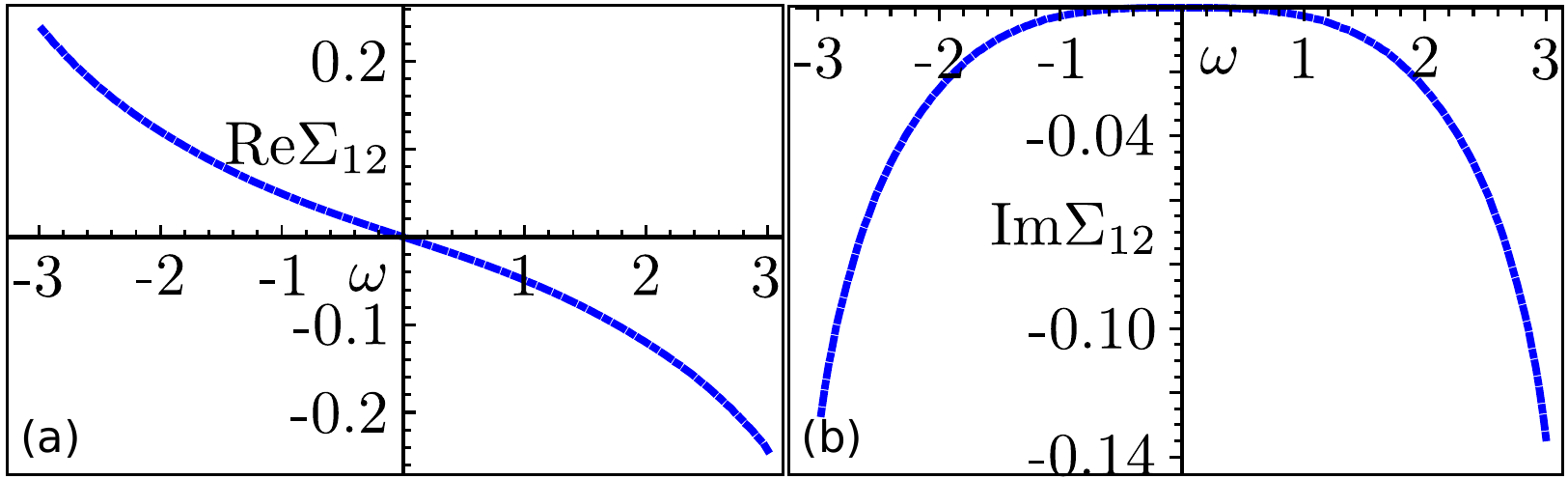}
\caption{The real (a) and imaginary (b) part of the off-diagonal self-energy $\Sigma_{12}$ (in the units of $t$) as a function of the frequency $\omega$ evaluated at the Dirac point at zero temperature, for the strength of the Hubbard repulsion $U = 0.6$ (in unit of $t$).}
\label{fig:MFL_Zero}
\end{figure}
%===========================================================================
Our goal, here, is to understand the modification of this Dirac structure as a function of $U$ in the weak coupling regime. We obtain the following second-order self-energy correction (see~\ref{sec:Apphubbard}) as
\begin{equation}\label{eq.11}
{\bm \Sigma}(\vk, \omega) = \begin{pmatrix}
\Sigma(\vk, \omega)  &  \Sigma_{12}(\vk,\omega) \\[3mm]
 \Sigma_{21}(\vk, \omega) & \Sigma(\vk, \omega)
\end{pmatrix},
\end{equation}
where ${\bm \Sigma}(\vk, \omega)$ denotes the $2\times 2$- self-energy matrix arising out of the spinor structure of the Dirac Hamiltonian. Individual components in the matrix entries are provided in ~\ref{sec:Apphubbard}. We begin with analyzing the renormalized spectral function $A(\vk, \omega)$ from Eq.~\ref{eq.11}. The bare spectral function for the non-interacting case with $U=0$ is shown in Fig.~\ref{fig:Fig4}a. The linear band crossing at the Dirac point $K$ is shown within the highlighted area in Fig.~\ref{fig:Fig4}a. In order to track the consequent band renormalization due to Hubbard repulsion, we analyze the spectral function $A(\vk,\omega)$. The latter is computed from the renormalized Green's function matrix $G(\vk,\omega)$ as $A(\vk, \omega) = -\text{Im} \Tr \big[G(\vk, \omega) \big]$.

The corresponding renormalized bands are shown in Fig.~\ref{fig:Fig4}b for various interaction strengths $U$ at zero temperature (only the highlighted portion of the spectrum in panel Fig.~\ref{fig:Fig4}a is considered here). On a closer look, we observe three separate lines as marked by filled circles. However, to clearly distinguish these three separated lines, we plot the spectral function $A(\mathbf{k}_P,\omega)$ at a fixed momentum $\mathbf{k}_P$ (marked by the dotted line in panel (b)) for different frequencies $\omega$ within an energy window from 0.8 eV to 1.2 eV (see Fig.~\ref{fig:Fig4}c). Indeed we observe three distinct peaks at three different $\omega$'s. Consequently, we infer that the bandwidth gradually decreases as a function of the coupling $U$, while keeping the Dirac point preserved. Hence, the Dirac velocity also decreases as a function of $U$.

\subsubsection{Analytical structure of $\bm{\Sigma}(\vk,\omega)$:}

%===========================================================================
\begin{figure}[t]
\centering
\includegraphics[width=1\linewidth]{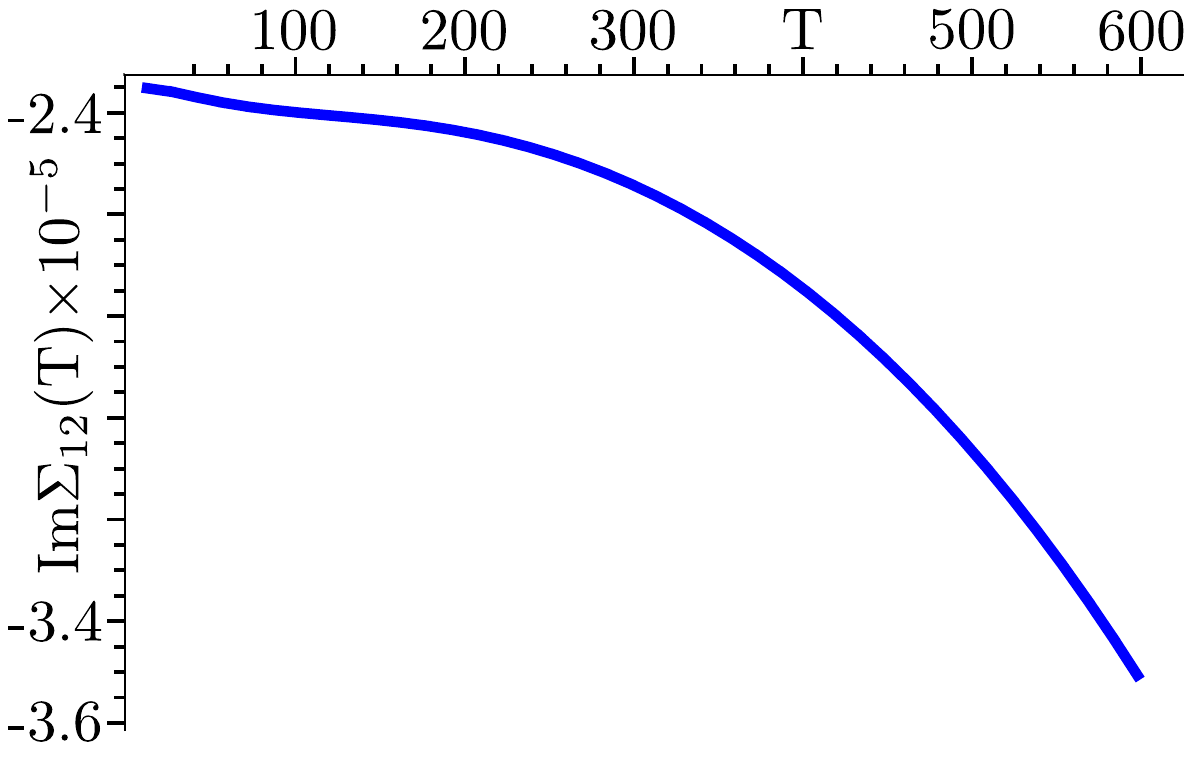}
\caption{The variation of the imaginary part of the off-diagonal self-energy $\Sigma_{12}(\text{K},\omega = 0)$ as a function of temperature. The Hubbard coupling is taken to be $U = 0.6t$.}
\label{fig:Temp_Predict}
\end{figure}
%===========================================================================
We now focus on analyzing the explicit frequency, momentum and temperature dependence of the self-energy components. We first analyze the self-energy components at zero temperature.  We observe that the diagonal component $\Sigma(\omega)$ at the Dirac point exhibits very weak frequency dependence; whereas the off-diagonal component $\Sigma_{12}(\omega)$ varies strongly with frequency (see Fig.~\ref{fig:MFL_Zero}a,b). To extract the frequency dependence of Im$\Sigma_{12}$, we depend on numerical fitting. We obtain the best fit to a numerical ansatz $A\cdot \omega^B$ with an exponent $B \sim 4$. We further notice that the other remaining parameter $A$ is estimated to be extremely small in magnitude. The origin of such a small magnitude is possibly connected to the vanishing quasiparticle density-of-states at the Dirac point. Using the Kramers-Kronig relation, we find that Re$\Sigma_{12}$ exhibits a $\omega^3$ frequency dependence.

 The imaginary part of the self-energy for a 2D Fermi liquid typically follows $\omega^2\log(\omega)$ frequency dependence~\cite{PhysRevB.75.121406,PhysRevB.89.205128}. Hence, our results dictate a deviation from the Fermi liquid picture. Considering the decrease of the band-width in Fig.~\ref{fig:Fig4}b along with the relativistic invariance of the quasiparticles around the Dirac point ($\omega = v_D k$), we predict the Dirac velocity renormalizes as
\begin{equation}\label{eq.Hubbard_Decrease}
\delta v_D \propto U^2 k^2.
\end{equation}

\subsubsection{Finite temperature:} So far, we discussed the renormalization of the Dirac spectrum at zero temperature. Now, we focus on the temperature evolution of the off-diagonal component $\Sigma_{12}$ at the Dirac point. The temperature evolution of  Im$\Sigma_{12}(\text{K},\omega=0)$ is shown in Fig.~\ref{fig:Temp_Predict}. Of course, it is very difficult to determine the form of temperature dependence just by examining our numerical results at a small temperature. However, at a larger temperature, our result closely resembles a quartic power-law dependence as in Fig.~\ref{fig:MFL_Fit}. Using the best fit, we predict a similar $T^4$ temperature dependence for the imaginary part of the self-energy.

%===========================================================================
\subsection{Short-range interactions: Dirac bosons \label{sec:dbose_sh}}

In the previous sections, we observed an interesting similarity between the interacting Dirac fermions. First, the low-energy fermionic excitations are still described by a modified Dirac-like Hamiltonian under the influence of the set of incorporated interactions. Secondly, long-range interactions typically lead to a logarithmic renormalization of the self-energy, whereas only power-law corrections are obtained for sufficiently localized interactions. We are, therefore, naturally tempted to analyze a few interacting bosonic systems to find if there are also such similarities in comparison to the fermionic case.

We begin with focusing on the two typical bosonic Dirac systems: (a) Cooper pairs that are described by an effective Bose-Hubbard model and (b) interacting magnons induced by a ferromagnetic Heisenberg model defined on a honeycomb lattice.

\subsubsection{\textbf{Cooper pairs: Extended Hubbard interaction} \label{sec:cooper_pair}}

Earlier theoretical work (see Ref.~\cite{Banerjee}), by some of the present authors, analyzed a physical realization for the granular superconducting network for Dirac-like bosons that arise from the collective excitations of the Cooper pairs. The corresponding Hamiltonian is described by a Bose-Hubbard model on a honeycomb lattice as
\begin{equation}\label{eq.20}
\mathcal{H}=-t\sum_{\langle ij \rangle}b^{\dagger}_{i}b_{j} + h.c.
+U\sum_i(n_{i}-n_{0})^2,
\end{equation}
where $b^{ \dagger}_{i}$ is a Cooper pair creation operator within a superconducting island at site $i$ for the honeycomb lattice. The former is defined as
\begin{equation}\label{eq.21}
b^{\dagger}_i = \sum_{\dr \in \mathcal{G}_i} c^{\dagger}_{\uparrow \dr} c^{\dagger}_{\downarrow \dr},
\end{equation}
where the summation in Eq.~\ref{eq.21} is over an individual grain $\mathcal{G}_i$ at site $i$ with $c^{\dagger}_{\uparrow \dr} (c^{\dagger}_{\downarrow \dr})$ being the fermionic creation operator with upward (downward) spin orientation.

The other parameters in Eq.~\ref{eq.20} are defined as follows: (i) $t$ is the hopping amplitude of the pairs between the sites, (ii) $U$ is an onsite potential, and (iii) $n_0$ denotes the mean-number of Cooper pairs in each grain. The collective excitations of this Cooper pair network produce an energy spectrum similar to the one shown in Fig.~\ref{fig:Spectrum}a. The two inequivalent modes participating in this excitation spectrum around the crossing point are effectively described by a Dirac-like Hamiltonian with a velocity $v_D = \frac{a\sqrt{3JU}}{4}$ with $J$ (=$n_0t$) representing the strength of the kinetic energy for these modes. However, when a nearest-neighbour interaction  $U'\sum_{\langle ij \rangle}n_{i}n_{j}$  is turned on, the Dirac velocity further modifies. A simple mean-field analysis predicts that the renormalized Dirac velocity decreases as
\begin{equation}\label{eq.22}
\delta v_D/v_D \approx -\frac{3U'}{2U},
\end{equation}
for $U' \lesssim U \ll J$ (see~\ref{sec:small_deri} for the details of the derivation of Eq.~\ref{eq.22}).  Hence, again, in this case, the algebraic structure of the Hamiltonian describing the states around the crossing point remains isomorphic to the non-interacting Dirac Hamiltonian, whereas the Dirac velocity acquires a quantitative downward modification.

\subsubsection{\textbf{Dirac magnons: Heisenberg interaction}}

Previously, we described that the short-range interactions lead to a decrease in the renormalized Dirac velocity of the interacting Dirac boson. However, one might naturally question the robustness of such a mean-field analysis. To answer this, we revisit our previous work~\cite{Saikat} on the ferromagnetic insulator CrBr$_3$, where Cr$^{3+}$ spins interact via Heisenberg exchange as
\begin{equation}\label{eq.23}
\mathcal H=-J\sum_{\langle ij \rangle}
{\textbf{S}}_{i}\cdot{\textbf{S}}_{j},
\end{equation}
where $J$ denotes the strength of the Heisenberg exchange interaction between the nearest-neighbor sites $\langle ij \rangle$ of a honeycomb lattice. Utilizing Holstein-Primakoff (HP) transformation, the spin operators $\textbf{S}_i$, in the above Hamiltonian, can be written in terms of the magnon creation and annihilation operators corresponding to the induced spin-wave excitations. The leading order terms in the Holstein-Primakoff expansion leads to the bosonic version of the Hamiltonian (see Eq.~\ref{eq.1}) already discussed in Sec.~\ref{sec:singleparticle}.
%===========================================================================
\begin{figure}[t]
\centering
\includegraphics[width=1\linewidth]{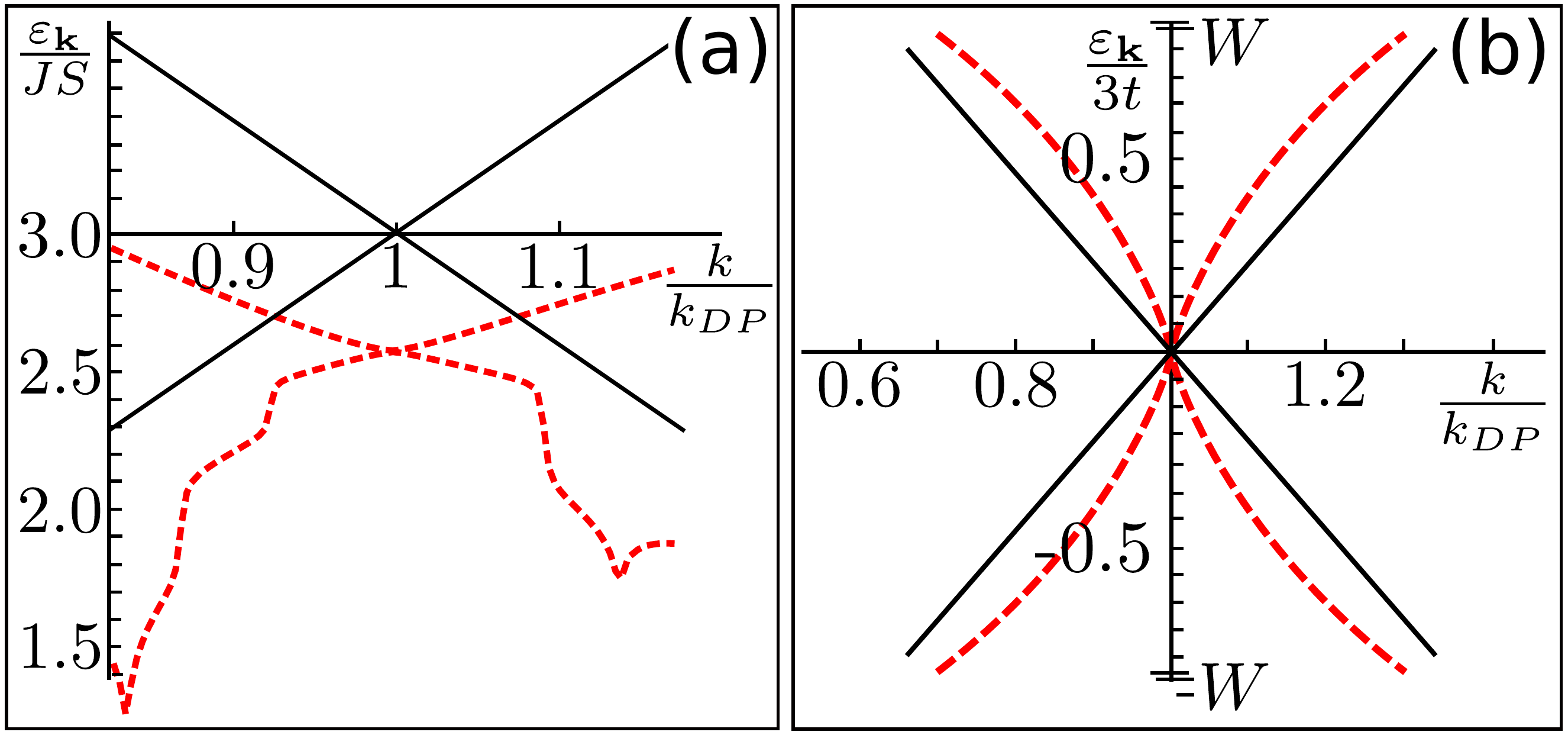}
\caption{Comparison of the renormalized band-structure of the interacting Dirac bosons (left panel) vs fermions (right panel) around the crossing point. (a) The conical Dirac spectrum (solid line) of the non-interacting spin-wave excitations in a honeycomb ferromagnet, arising from the Holstein-Primakoff transformation (HPT) in Eq.~\ref{eq.23}. The leading higher order terms of the HPT leads to a short-range interaction resulting in a renormalized spectrum (dashed line). Along with a finite shift in the Dirac point energy the Dirac velocity decreases. Reproduced from Ref.~\cite{Saikat}. (b) The bare conical spectrum (solid line) of the Dirac fermions renormalizes due to the long-range interaction leading to an increased Dirac velocity as illustrated in the modified spectrum (dashed line).}
\label{fig:renorm_spectrum}
\end{figure}
%===========================================================================

%===========================================================================
\begin{figure}[b]
\centering
\includegraphics[width=1\linewidth]{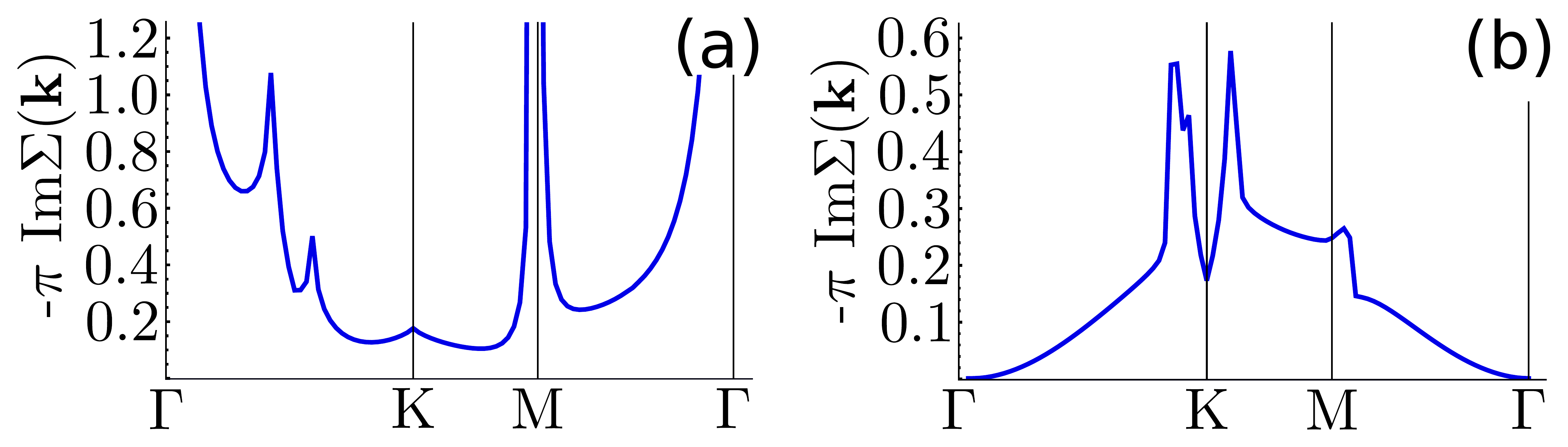}
\caption{Scattering rates of the up (a) and down (b) bands of the spin-wave excitation spectrum (see Fig.~\ref{fig:Spectrum}a), due to the leading higher order terms of the Holstein-Primakoff transformation of the Heisenberg model in Eq.~\ref{eq.23}. The rates (in the units of $JS$) are shown along the high-symmetry points in the Brillouin zone of a honeycomb lattice. Reproduced from Ref.~\cite{Saikat}.}
\label{fig:Fig3}
\end{figure}
%===========================================================================

The subleading higher-order terms, essentially yield an effective model of the Dirac magnons interacting via short-range interactions, almost similar to nearest-neighbor Hubbard interaction discussed in the previous section. The resulting renormalization of the magnons has been studied in detail within a second-order perturbation theory in our previous work~\cite{Saikat}. For the purpose of the recent paper, here we skip the corresponding explicit derivations and recollect the primary results: (a) the renormalized spectrum around the Dirac crossing point (see Fig.~\ref{fig:renorm_spectrum}a), and (b) induced decay rate of the spin-wave excitations (see Fig.~\ref{fig:Fig3}).

 We observe that the renormalized Dirac velocity, in this case, decreases (the slope of the modified cone (dashed line) is smaller than the slope of bare Dirac cone (solid line) in Fig.~\ref{fig:renorm_spectrum}a). We further observe that the magnonic excitations around the Dirac point acquire momentum-dependent decay rates as shown in Fig.~\ref{fig:Fig3}. Both the real and imaginary parts of the self-energy contain a $T^2$ temperature dependence.

\subsection{Long-range interactions: Dirac bosons \label{coulomb_bose}}

So far, we discussed the renormalization behavior of various interacting Dirac systems. The results derived or collected from the existing literature, throughout Sec.~\ref{sc_coulomb_fermi} \--- Sec.~\ref{sec:dbose_sh} already point to an interesting similarity between the fermionic and bosonic Dirac matter: short-range interactions qualitatively leads to a decreased Dirac velocity with power-law self-energy corrections. In contrast, long-range interactions yield an increased Dirac velocity for fermions with non-polynomial (logarithmic) self-energy renormalization. Hence, one may ask whether a similar behavior is replicated in case for Dirac bosons coupled via a long-range interaction.

In this section, we consider a model for charged bosonic quasiparticles interacting via the long-range Coulomb repulsion on a honeycomb lattice. The corresponding Hamiltonian is written as
\begin{equation}\label{eq.24}
H = -t\sum_{\langle ij \rangle} b^{\dagger}_i b_j + 3t\sum_i b^{\dagger}_ib_i + \frac{{e^*}^2}{\epsilon_0}\sum_{ij} \frac{n_in_j}{|\dr_i - \dr_j|},
\end{equation}
 where $b^{\dagger}_i$ denotes the creation operator for the bosonic excitation with charge $e^*$, $t$ is hopping amplitude between nearest-neighbor sites on the honeycomb lattice and the last term denotes the Coulomb interaction between the two different sites $i,j$. Motivated by the free spin-wave theory on a honeycomb lattice (see Heisenberg model discussion in Sec.~\ref{sec:dbose_sh}), we  introduced the onsite energy $3t$ in the model in Eq.~\ref{eq.24} to constraint positive definite spectrum for the bosons.

To analyze the effect of the Coulomb interaction on the Dirac spectrum, the Hamiltonian (Eq.~\ref{eq.24}) is rewritten in the diagonal basis. We focus on a subset of the scattering processes which are relevant at low temperature (see~\ref{sec:Appcoulomb} for the details). We notice that there are eight different scattering processes that take part in the Coulomb interaction in the diagonal basis. We assume that one of the interacting bosons ($d_{\vk'}$) is always excited by the thermal energy, and due to low temperature will be occupied near the bottom of the band-structure as illustrated in Fig.~\ref{fig:Spectrum}a. The other bosonic excitations can, in principle, be created anywhere in the band-structure using some external source (\textit{viz.} in case of magnons, this external source could be inelastic neutron scattering). Under this approximation, we retain only three of all the scattering channels (see Eq.\ref{appeq.27}) as
\begin{align}\label{eq.25}
&\mathcal{H}_{\text{int}}  = \sum_{\vq,\vk,\vk'} V_\vq \Big[ 2\cos\Phi u^{\dagger}_{\vk-\vq}d^{\dagger}_{\vk'+\vq}u_{\vk}d_{\vk'} +  \\
& i\sin\Phi d^{\dagger}_{\vk-\vq}d^{\dagger}_{\vk'+\vq}u_\vk d_{\vk'} + \scalebox{0.93}[1]{$\left(\frac{2\cos\Phi-\cos\Psi}{4}\right)$} d^{\dagger}_{\vk-\vq}d^{\dagger}_{\vk'+\vq}d_\vk d_{\vk'} \Big], \nonumber
\end{align}
where the angles $\Phi, \&\, \Psi$ are defined in \ref{sec:Appcoulomb}. Assuming the strength of the Coulomb interaction is small compared to kinetic energy (${e^*}^2/\epsilon_0a_s < t$, where $a_s$ is the Born scattering length), we set up a second-order perturbation theory. The self-energy for the respective channels are derived in~\ref{sec:Appcoulomb} as
\begin{equation}\label{eq.26}
\Sigma(\vk) \cong \sum_{\vk',\vq}\frac{|V_{\vq,\vk}|^2 b(\varepsilon^d_{\vk'})}{\varepsilon_\vk - \varepsilon_{\vk-\vq}-\varepsilon_{\vq}+i\delta}.
\end{equation}
We notice that the temperature dependence of the self-energy in all the three channels is governed by the factor $\sum_{\vk'}b(\varepsilon^d_{\vk'})$; where the dispersion for the thermally excited bosons is approximated as $\varepsilon^d_{\vk'} \approx \frac{3}{4}t|\vk'|^2$. To evaluate this integral, we introduce an infrared cut-off $k_0 \sim L^{-1}$, where $L$ is the linear system size. The integral is consequently evaluated as
\begin{align}\label{eq.27}
& 2\pi\int_{k_0}^{\infty} \frac{k'd k'}{e^{ \frac{3\beta t}{4}k'^2}-1}  \approx  -\frac{4\pi\log (3\beta  t/4N^2)}{ 3\beta  t}
\end{align}
where $L=Na$ and $\beta = 1/T$ in the unit of Boltzmann constant $k_B = 1$. At this point, we would like to mention that the possibility of the formation of a Bose-Einstein condensate is ignored by assuming that the relevant temperature is always larger than the corresponding condensation temperature (the condensation temperature for parabolic bosons is absolute zero in the two-dimensions). Consequently, we obtain
\begin{equation}\label{eq.28}
\Sigma(\vk) \approx \frac{4 T \pi}{ 3t}\log \Big[\frac{4TN^2}{3t}\Big]\sum_{\vq}\frac{|V_{\vq,\vk}|^2}{\varepsilon_\vk - \varepsilon_{\vk-\vq}-\varepsilon_{\vq}+i\delta}.
\end{equation}
The diverging logarithmic pre-factor in the bosonic case is in stark contrast to the logarithmic divergence for the Dirac fermions case (see Eq.~\ref{eq.4}). In the latter case, it is associated with an ultraviolet energy scale ($\Lambda$), whereas for the bosonic system, we have an infrared energy scale ($N$) which dictates the overall size of the underlying system. The momentum-dependent renormalization is consequently governed by the remaining integral in Eq.~\ref{eq.28}. The relevant analysis of the momentum-dependence of the self-energy is provided in~\ref{sec:Correction}. The renormalized spectrum is shown in Fig.~\ref{fig:App_Fig6}. We notice that the Coulomb repulsion does not lead to any gap opening at the Dirac point. Instead, the Dirac velocity is renormalized upward.

\section{Dirac materials in other dimensions \label{sec:db_dimension}}

In this section, we briefly discuss a few Dirac systems in one and three dimensions and outline the role of interactions. We start from the most common example of one dimensional (1D) Dirac excitations \--- the solutions of the Su-Schrieffer-Heeger (SSH) model~\cite{RevModPhys.60.781} which describes spinless fermions on a dimerized 1D lattice with staggered hopping amplitudes. In this case, the effective Hamiltonian around the edge of the Brillouin zone describes massive Dirac fermions. The mass term disappears when the two in-equivalent hopping amplitudes become identical, and consequently leads to nodal Dirac fermions. It is natural to expect that a bosonic analog of SSH model would lead to similar Dirac bosons at the edge of the one-dimensional Brillouin zone. Indeed, there have been recent theoretical predictions about such collective bosonic excitations on plasmons~\cite{PhysRevB.95.125426}, magnons~\cite{PhysRevB.98.224409}, polaritons~\cite{PhysRevLett.123.217401}, etc. While the primary focus on these one dimensional Dirac systems has been on the topological aspects of the corresponding system, far less investigation is done on the electronic interaction effects. Previous studies of short-range interactions on fermionic SSH model based on the renormalization group analysis~\cite{PhysRevB.27.1680} indicates that the low-energy Dirac-like behavior belongs to a two-component Gross-Neveu class, in a similar way to the two-dimensional Dirac fermions. Based on our analysis of two dimensional systems in this paper, we predict that the Dirac velocity will also renormalize downward in the 1D case for short-range interactions. In this spirit, the renormalization should be upward for any long-range interactions. 

Renormalization effects of three-dimensional (3D) Weyl-Dirac fermions were studied in detail in Ref.~\cite{PhysRevB.92.115101}, which found that the Dirac velocity is renormalized upwards in the presence of Coulomb interaction. This result is consistent within our predictions that long-range interaction generically leads to an increase in the Dirac velocity. Consistently, the short-range interactions remain marginally irrelevant~\cite{PhysRevB.96.155112} and lead to a decrease in the Dirac velocity. To investigate whether this feature persists in the 3D bosonic Dirac systems, we need to incorporate explicit model calculation for various bosonic systems, which is left for future work.

%============================================================================
\section{Discussion and conclusion \label{sec:discussion}}

In this paper, we present a systematic study of the renormalization effects in Dirac materials for various interactions in the weak coupling regime. The results recorded in Table~\ref{Summary}, give a unified summary of the effect of the interactions on the Dirac quasiparticles for the models studied/discussed in this paper. The form of the quasiparticle interactions qualitatively affects the velocity renormalization \--- contact or short-range interactions generically lead to a decrease in the Dirac velocity, whereas long-range interactions (both screened and unscreened) typically leads to an increase. The quasiparticle statistics \--- i.e. whether we are dealing with bosons or fermions \--- does affect the power laws fixing the temperature-dependence of the self-energy corrections for short-range interactions, but does not influence the logarithmic behavior for the long-range Coulomb repulsion.

In addition to the discussion on the overall renormalization of the Dirac velocity, we found non-Fermi liquid behavior for interacting Dirac fermions. Our analysis in Sec.~\ref{sec:hubbard_int} on the Hubbard model on honeycomb lattice leads to a quartic frequency-dependent Im$\Sigma(\omega)$, in comparison to the conventional $\omega^2\log \omega$ structure for the two-dimensional Fermi liquid. We furthermore mention that the primary purpose of the bosonic analysis (in Sec.~\ref{coulomb_bose} and \ref{sec:Correction}) is to capture the overall trend for the renormalized band structure around the Dirac point without aiming for a precise quantitative estimate.

A closer look at Table~\ref{Summary} implies that the power-law exponents in the self-energy for the short-range interactions crucially depend on the specific model. However, the common feature in all these situations is the overall decrease of the Dirac velocity governed by a polynomial type correction for both the real and imaginary part of the self-energy. On the other hand, the long-range interactions in both the fermionic and bosonic cases, yield logarithmic self-energy corrections. In both these cases, the renormalized Dirac velocity consists of divergences. Yet, the nature of such divergences differs between the two statistical species \--- fermions have an ultraviolet divergence compared to an infrared singularity in the bosonic case. We admit that there are definitely certain intrinsic differences between fermionic and bosonic Dirac materials in the interacting regime. At the same time, the qualitative similarity of the results based on Table~\ref{Summary} is naturally reflected based on (a) slowing down or speeding up the Dirac quasiparticles and (b) the appearance of polynomial or non-polynomial correction in the self-energy.

It is important to note that many of the fermionic renormalization features analyzed in this paper are also verified in various experimental observations, \textit{viz.}  Quantum Oscillations experiments in suspended graphene~\cite{Elias} can be used to extract the non-linear momentum dependence of the Dirac velocity (Sec.~\ref{coulomb_fermi},\ref{sc_coulomb_fermi}). On the other hand, the diamagnetic susceptibility in two dimensional Dirac fermionic systems typically diverges at zero temperature~\cite{PhysRevB.75.115123,ANDO2007213}. However, the unscreened Coulomb repulsion leads to a strong enhancement of this feature and removes the divergences~\cite{PhysRevLett.99.226803}. Hence, experimental observation of the diamagnetic susceptibility can, in principle, display the features of the non-linear renormalization for long-range interactions in the two-dimensional fermionic systems. We notice that the emergence of non-Fermi liquid behavior for the Dirac fermions under the influence of long-range interaction is responsible for the dramatic changes in the corresponding physical observables.

 However, in the case of the short-range interactions in Dirac fermions, such non-Fermi liquid characteristics arise only to the sub-leading correction in the self-energy. Hence, it becomes a challenging task to observe the fingerprints of such corrections (Sec.~\ref{sec:hubbard_int}) in a physical quantity. Precise measurements of various quantities \textit{viz.} magnetoresistance, quantum Hall-effect, etc may show deviations from the Fermi liquid predictions which can be associated with the renormalization corrections described earlier.

  For the bosonic systems, the experimental observation of the features of the spectral renormalization around the high-energy crossing point is complicated, as the low-energy bosons near the ground state would naturally couple to any experimental probe. Additionally, the quasiparticle excitations are neutral in most of the conventional bosonic systems.  We expect that the measurement of various thermal quantities may show the effects of the renormalization of the conical Dirac spectrum.

To conclude, our results show how the quasiparticles in Dirac materials are modified by various interactions. While our results are based on a specific subset of the diverse realizations of Dirac materials, a universal description of the same for both fermionic and bosonic quasiparticles remains an interesting possibility.

%============================================================================
\section{Acknowledgments}
S.B. acknowledges special thanks to Arno P. Kampf. We are grateful to D. Vollhardt, M. Kollar, W. A. Atkinson, F.F. Assaad, D.N. Basov, L. Chioncel, P. Hofmann and S.S.  Pershoguba for discussions. S. B. acknowledges support by the Deutsche
Forschungsgemeinschaft through TRR 80. H. {\AA}. acknowledges the Knut and Alice Wallenberg Foundation for financial support (Grant No. KAW-2013.0020).  Work at KTH was supported by ERC DM 321031,  the KAW-2013.0096  and the  VILLUM FONDEN via the Centre of Excellence for Dirac Materials (Grant No. 11744).

\appendix
%===========================================================================
\section{Derivations of the effective Dirac Hamiltonian \label{sec:small_deri}}

In this section, we provide the details of the derivation of the effective Dirac Hamiltonian starting from the tight-binding description on the honeycomb lattice. The fermionic Hamiltonian in Eq.~\ref{eq.1} can be rewritten in momentum space as
\begin{equation}\label{rappeq.1}
H = - t\sum_\vk \Big[ \gamma_\vk c^{\dagger}_{a\vk} c_{b\vk} + \gamma^{\ast}_\vk c^{\dagger}_{b\vk} c_{a\vk} \Big],
\end{equation}
where the chemical potential $\mu$ is assumed to be zero, $c^{\dagger}_{\alpha\vk}$ creates an electron on the $\alpha$ sublattice site and $\gamma_\vk$ is defined as
\begin{equation}\label{rappeq.2}
\gamma_\vk = \sum_{i=1}^3 e^{i\vk \cdot \bm{\delta}_i},
\end{equation}
where $\delta_i$ are the three nearest-neighbor vectors as shown in Fig.~\ref{fig:App_Fig2}a
\begin{equation}
\delta_1 = \frac12(1,\sqrt{3}), \; \delta_2 = \frac12(1,-\sqrt{3}), \delta_3 = \frac12(-1,0),
\end{equation}
in the unit of the lattice constant $a$. Please note that the irrelevant spin degrees of freedom has been neglected. The Hamiltonian in Eq.~\ref{rappeq.1} can now be rewritten in the matrix form as
\begin{equation}\label{rappeq.3}
H = \sum_\vk \Psi^{\dagger}_\vk \mathbf{H}_\vk \Psi_\vk, \; \mathbf{H}_\vk =-t\begin{pmatrix}
0 & \gamma_\vk \\
\gamma^{\ast}_\vk & 0
\end{pmatrix},
\end{equation}
where $\Psi_\vk$ is a spinor composed of the two sublattice operators as $\Psi^{\dagger}_\vk = (c^{\dagger}_{a\vk}, c^{\dagger}_{b\vk})$. Expanding the $2\times2$-Hamiltonian in Eq.~\ref{rappeq.3} near the Dirac point K = $(\frac{2\pi}{3}, \frac{2\pi}{3\sqrt{3}})$ we obtain
\begin{equation}\label{rappeq.4}
\mathbf{H}_{\text{K} + \vk} =- \frac{3t}{2}\begin{pmatrix}
0 & k_x - ik_y \\
k_x +i k_y & 0
\end{pmatrix}.
\end{equation}
Rewriting the effective Hamiltonian with Pauli matrices we obtain the effective Hamiltonian in Eq.~\ref{eq.2} with the Dirac velocity $v_D = \frac{3t}{2}$.

A bosonic analogue of the effective Hamiltonian resulting from the Heisenberg interaction in Eq.~\ref{eq.23}, is obtained by rewriting the spin operators $S_i$ in terms of magnon operators. The latter is obtained by Holstein-Primakoff transformation (HPT)~\cite{Saikat} as
\begin{align}\label{rappeq.5}
S^x_i + i S^y_i & = \sqrt{2S} \left( a_i - \frac{a^{\dagger}_i a_ia_i}{4S}  \right) + \mathcal{O}(1/S^{\frac32}), \nonumber \\
S^x_i - i S^y_i & = \sqrt{2S} \left( a^{\dagger}_i - \frac{a^{\dagger}_i a^{\dagger}_ia_i}{4S}  \right) + \mathcal{O}(1/S^{\frac32}), \nonumber \\
S^z_i & =  S - a^{\dagger}_ia_i,
\end{align}
where $S$ is the total spin. We substitute the leading order terms of HPT in Eq.~\ref{eq.23} to obtain the effective bosonic Hamiltonian in the momentum-space as
\begin{equation}\label{rappeq.6}
H = \sum_\vk \Psi^{\dagger}_\vk \mathbf{H}_\vk \Psi_\vk, \; \mathbf{H}_\vk =JS\begin{pmatrix}
3 & -\gamma_\vk \\
-\gamma^{\ast}_\vk & 3
\end{pmatrix},
\end{equation}
where the spinor $\Psi_\vk$ is defined in terms of the two inequivalent boson operators on the two sublattice sites for an honeycomb lattice. $\gamma_\vk$ is the same as defined earlier in Eq.~\ref{rappeq.2}. The diagonal terms $3JS$ arises from the $S^z_iS^z_j$ coupling in the Heisenberg interaction of Eq.~\ref{eq.23}. In contrast to the fermionic case, we keep the entire structure in the full Brillouin zone. The band structure from the Hamiltonian in Eq.~\ref{rappeq.6} is shown in Fig.~\ref{fig:Spectrum}a.

Finally, we provide the details of the mean-field result in Eq.~\ref{eq.22}. We start from an extended Bose-Hubbard model defined on a honeycomb lattice as
\begin{equation}\label{rappeq.8}
\mathcal{H} = -t \sum_{\langle ij\rangle} b^{\dagger}_ib_j + U\sum_{i}(n_i-n_0)^2+ U'\sum_{\langle ij\rangle} n_in_j,
\end{equation}
where $b^{\dagger}_i$ is the bosonic quasiparticle creation operator and $U,U'$ are the strengths of the onsite and nearest-neighbor Hubbard interactions, respectively. We further assume that there are $n_0$ mean number of bosons per grains. Now we translate the bosonic operator to their charge-density representation as $b^{\dagger \text{A}}_i = \sqrt{n_i^{\text{A}}}e^{i\theta_i^{\text{A}}}$ and $b^{\dagger \text{B}}_i = \sqrt{n_i^{\text{B}}}e^{i\theta_i^{\text{B}}}$, where $\text{A/B}$ corresponds to the two sublattice sites. Assuming the bosons are now Cooper pairs on a granular superconductor as described, in Sec.~\ref{sec:cooper_pair}, we focus on the low-energy effective theory of the fluctuations of the phase $\theta_i$'s. For small $U$ and $U'$, we focus on the small phase fluctuations around mean boson density $n_0$ and consequently obtain the following non-interacting Hamiltonian as
\begin{align}\label{rappeq.9}
\mathcal{H} = \sum_{\vk} \Big[ & 3J\delta\theta^{\text{A}}_\vk \delta\theta^{\text{A}}_{-\vk} + 3J\delta\theta^{\text{B}}_\vk \delta\theta^{\text{B}}_{-\vk} - \gamma_\vk \delta\theta^{\text{A}}_\vk \delta\theta^{\text{B}}_{-\vk} \nonumber \\
& -\gamma^{\ast}_\vk \delta\theta^{\text{B}}_\vk \delta\theta^{\text{A}}_{-\vk} + U \delta n^{\text{A}}_\vk \delta n^{\text{A}}_{-\vk} + U \delta n^{\text{B}}_\vk \delta n^{\text{B}}_{-\vk} \nonumber \\
& + \frac{U'}{2}\gamma_\vk \delta n^{\text{A}}_\vk \delta n^{\text{B}}_{-\vk} + \frac{U'}{2} \gamma^{\ast}_{\vk} \delta n^{\text{A}}_\vk \delta n^{\text{A}}_{-\vk} \Big],
\end{align}
where the differential phases and densities are small fluctuations around the ground state $(n_0, \theta_0)$ and $J = n_0t$. We now solve for the normal mode frequencies from the Hamiltonian in Eq.~\ref{rappeq.9} and obtain the frequencies for the two modes as~\cite{Banerjee}
\begin{equation}
\omega_{1,2}(\vk) = J \left(3U \mp \frac{3U'}{2}|\gamma_\vk| \pm U |\gamma_\vk| - \frac{U'}{2}|\gamma_\vk|^2 \right).
\end{equation}
The modification of Dirac velocity can now be easily extracted as in Eq.~\ref{eq.22} by taking the limit $U'$ to zero.

%===========================================================================
\section{Coulomb repulsion at finite temperature \label{sec:AppHFT}}

In Sec.~\ref{coulomb_fermi}, we discussed the disappearance of the logarithmic divergence of the Dirac velocity at a finite temperature as in Eq.~\ref{eq.4}. Here, we outline the details of the derivation of this result. A generic two-body interaction $V(\dr,\dr')$ can be formulated as
\begin{equation}\label{appeq.1}
H = \int d\dr d\dr' \psi^{\dagger}_{\dr} \psi^{\dagger}_{\dr'} V(\dr,\dr') \psi_{\dr'}\psi_{\dr},
\end{equation}
where $\psi_{\dr}$ is a fermionic field. Eq.~\ref{appeq.1}, which contains four fermionic fields, can be recasted in terms of two-fermionic fields by introducing an auxiliary bosonic field $\phi_{\dr}$ as
\begin{equation}\label{appeq.2}
H = \int d\dr \phi_{\dr} \psi^{\dagger}_{\dr} \psi_{\dr} + \int d\dr d\dr' \frac{\phi^{\dagger}_{\dr} \phi_{\dr'}}{V(\dr,\dr')}.
\end{equation}
Eq.~\ref{appeq.2} is obtained from Eq.~\ref{appeq.1} using the Hubbard-Stratonovich transformation. It is important to mention that the original Hamiltonian in Eq.~\ref{appeq.1} can be recovered by integrating out the
%===========================================================================
\begin{figure}[b]
\centering
\includegraphics[width=0.6\linewidth]{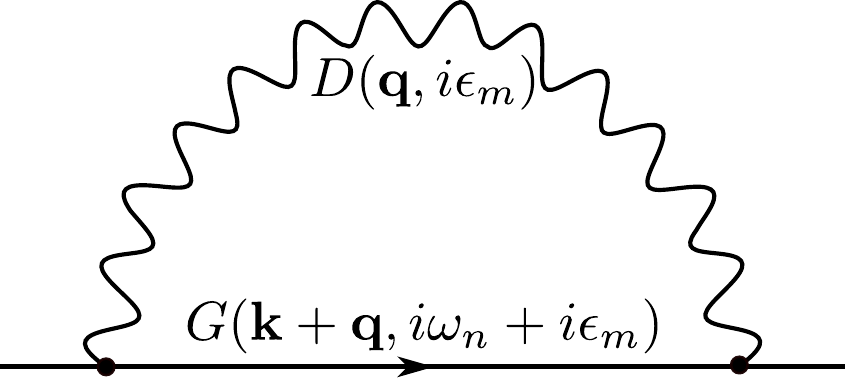}
\caption{Hartree-Fock self-energy diagram that leads to the self-energy expression in Eq.~\ref{appeq.5} and the consequent temperature-dependent renormalization of the Dirac velocity in Eq.~\ref{eq.4}.}
\label{fig:App_Fig1}
\end{figure}
%===========================================================================
auxiliary bosonic field $\phi_{\dr}$, which can be thought of as an instantaneous mediator of the exchange interaction $V(\dr,\dr')$ between the two fermionic fields. The dynamics of $\phi_{\dr}$ is governed by its Green's function
\begin{equation}\label{appeq.3}
\Braket{\phi^{\dagger}_{\dr,t} \phi_{\dr',t'}} = V(\dr,\dr'),\; D(\vq, i\epsilon_m) = V(\vq),
\end{equation}
where $D(\vq, i\epsilon_m)$ is the Green's function in momentum space, with $V(\vq)$ being the Fourier transform of $V(\dr,\dr')$ with $\epsilon_m$ being the bosonic Matsubara frequency. We now consider the non-interacting fermionic Green's function. Following the Hamiltonian in Eq.~\ref{eq.2}, we obtain the latter as
\begin{equation}\label{appeq.4}
G(\vk,i\omega_n)= \frac{1}{i\omega_n-v_D {\bm \sigma} \cdot \vk},
\end{equation}
where $\omega_n$ is the fermionic Matsubara frequency.

Next, we consider the lowest order (in the dimensionless coupling $\alpha$) self-energy correction to the Green's function in Eq.~\ref{appeq.4} as (see Fig.\ref{fig:App_Fig1})
\begin{equation}\label{appeq.5}
\Sigma (\vk,i\omega_n)  = -\frac{1}{\beta}\sum_{\omega_n,\vq} G(\vk+\vq,i\omega_n+i\epsilon_m)D(\vq,i\epsilon_m),
\end{equation}
where $\beta= \frac{1}{k_B T}$ and $k_B$ is the Boltzmann constant. Relabeling $\omega'_n = \epsilon_m+\omega_n$ and $\vq \to \vq-\vk$, we simplify the integrand as
\begin{equation*}
G(\vq,i\omega'_n)D(\vq-\vk, i\omega'_n-i\omega_n)
= \frac{i\omega'_n + v_D \bm{\sigma} \cdot \vq}{(i\omega'_n)^2-v_D^2q^2} \frac{e^2}{\epsilon_0|\vk-\vq|},
\end{equation*}
where $V(\dr,\dr')$ is taken to be the Coulomb interaction. Now, we compute the frequency summation in Eq.~\ref{appeq.5} as
\begin{align}\label{appeq.6}
& \frac{1}{\beta}  \sum_{\omega'_n}\frac{i\omega'_n + v_D \bm{\sigma} \cdot \vq}{(i\omega'_n)^2-v_D^2q^2}  \nonumber \\
& = \frac{1}{\beta}\sum_{i\omega'_n} \frac{1}{2}\Big[ \frac{1}{i\omega'_n - v_D q} + \frac{1}{i\omega'_n + v_D q} \Big] + \frac{v_D {\bm \sigma} \cdot \vq}{(i\omega'_n)^2-v_D^2q^2}  \nonumber \\
& = -\frac{1}{2}\Big[f(v_Dq) + f(-v_Dq) \Big] - \frac{1}{2}\Big[ \frac{e^{\frac{\beta v_D q}{2}}-e^{-\frac{\beta v_D q}{2}}}{e^{\frac{\beta v_D q}{2}}+e^{-\frac{\beta v_D q}{2}}} \Big] \frac{\bm{\sigma} \cdot \vq}{q} \nonumber \\
& = -\frac{1}{2} - \frac{1}{2}\tanh\left( \frac{\beta v_D q}{2} \right)  \frac{\bm{\sigma} \cdot \vq}{q},
\end{align}
where $f(\varepsilon)$ is the Fermi-Dirac distribution function. Hence, the self-energy $\Sigma(\vk, i\omega_n)$ in Eq.~\ref{appeq.5} becomes
\begin{align}\label{appeq.7}
\Sigma(\vk) = & \sum_{\vq} \Bigg[1+\tanh\left( \frac{\beta v_D q}{2} \right) \frac{\bm{\sigma} \cdot \vq}{q} \Bigg]  \frac{e^2}{2\epsilon_0|\vk-\vq|},
\end{align}
where the irrelevant frequency dependence in the self-energy has been dropped. The first term in the bracket in Eq.~\ref{appeq.7} corresponds to a constant shift in energy. Hence, we focus on the second term in the ultraviolet (large momentum behavior of the integrand in Eq.~\ref{appeq.7}) the regime as (in the unit of $e^2/2\epsilon_0$)
\begin{align}\label{appeq.8}
\Sigma(\vk) \propto \frac{k {\bm \sigma}}{2}  \cdot \int \lambda d\lambda d\theta  \frac{(\cos \theta, \sin \theta) \tanh \left( \frac{\beta v_D k \lambda}{2} \right)}{\sqrt{1+ \lambda^2 -2\lambda \cos(\gamma-\theta)}},
\end{align}
where the variable $\lambda$ is defined as $\lambda = q/k$. The exchange wavevector is written in polar coordinate as $\vq = q(\cos\theta, \sin\theta)$ and the angle between $\vk$ and $\vq$ is $\gamma - \theta$. For large $q \to \Lambda$ (where $\Lambda \sim 2\pi/a$ is the ultraviolet cutoff for the momentum in honeycomb lattice), the integrand in Eq.~\ref{appeq.8} reduces to
\begin{align}\label{appeq.9}
\Sigma(\vk) & \propto \frac{k {\bm \sigma}}{2}  \cdot \int^{\frac{\Lambda}{k}} d\lambda d\theta \Big[ (\cos \theta, \sin \theta) \cos(\gamma-\theta) \cdot \nonumber \\
 &  \qquad \qquad \qquad \qquad\tanh \left( \frac{\beta v_D k \lambda}{2} \right) \Big]\nonumber \\
& = \frac{{\bm \sigma}}{2}  \cdot \int^{\frac{\Lambda}{k}} \frac{d\lambda}{\lambda} \tanh \left(\frac{\beta v_D k \lambda}{2} \right) (k\cos \gamma, k\sin \gamma) \nonumber \\
& \xRightarrow[\Lambda \to \infty]{} \frac{{\bm \sigma} \cdot \vk}{2}  \log \left( \frac{v_D\Lambda}{2 k_B T}\right).
\end{align}
Hence, we obtain the self-energy correction (see Eq.~\ref{eq.4}) as $\Sigma(\vk) = \frac{e^2}{2\epsilon_0} \log \left( \frac{v_D\Lambda}{2 k_B T}\right) {\bm \sigma} \cdot \vk$.

To provide a pedagogical description, we finally provide a brief outline of the derivation leading to the well-known logarithmic renormalization of the Dirac velocity at zero temperature, as in Eq.~\ref{eq.3}. In this case, we compute the Hartree-Fock correction to the quasiparticle energy as
\begin{equation}\label{rappeq.7}
\delta E_\vk \approx \pm \int d\vk' \frac{e^2}{\epsilon_0|\vk-\vk'|} \left(\frac12 \pm  \frac{\vk \cdot\vk'}{|\vk||\vk'|}\right),
\end{equation}
where $\pm$ sign corresponds to the electrons and holes, respectively and the second term originates from the overlap of the eigenfunctions from the Hamiltonian in Eq.~\ref{rappeq.4}. Performing a similar large momentum expansion we find the divergent part of the integral in Eq.~\ref{rappeq.7} is proportional to $\pm \log \frac{\Lambda}{k}$.

%===========================================================================
\section{Hartree-Fock correction: Screened Coulomb interaction \label{sec:AppSY}}

In this appendix, we outline the derivation of the renormalization that led to Eq.~\ref{eq.8} and Eq.\ref{eq.9} in Sec.~\ref{sc_coulomb_fermi}. We first begin with the Fourier transformation of the screened Coulomb interaction $V_{SC}(\dr)$ (see Eq.~\ref{eq.7}) as
\begin{equation}\label{appeq.10}
V_{SC}(\vq) = \frac{e^2}{\epsilon_0} \frac{1}{q+q_{TF}}.
\end{equation}
Following the similar Hubbard-Stratonovich transformation as in Appendix~\ref{sec:AppHFT}, we obtain the zero temperature Green's function for auxiliary field $\phi_{\dr}$ as
\begin{equation}\label{appeq.11}
D(\vq,\epsilon) = \frac{e^2}{\epsilon_0} \frac{1}{q+q_{TF}}.
\end{equation}
Performing a Hartree-Fock analysis (replace the $\phi$-propagator in Fig.~\ref{fig:App_Fig1} by Eq.~{\ref{appeq.11}}, we obtain the lowest-order (in the dimensionless coupling $\alpha$) self-energy correction (in the unit of $e^2/\epsilon_0$) as
\begin{equation}\label{appeq.12}
\Sigma(\vk) \propto {\bm \sigma} \cdot \int q dq d\theta \frac{(\cos \theta, \sin \theta)}{|\vq-\vk|+q_{TF}},
\end{equation}
where we wrote the exchange wavevector in polar coordinate as $\vq = q(\cos\theta, \sin\theta)$. Choosing $\gamma-\theta$ as the angle between the two wavevectors $\vk$ and $\vq$, we rewrite Eq.~\ref{appeq.12} as
\begin{equation}\label{appeq.13}
\scalebox{1.05}[1]{$\Sigma(\vk) \propto k' {\bm \sigma} \cdot \int \frac{\lambda d\lambda (\cos \theta, \sin \theta)  d\theta}{\sqrt{\lambda^2-2\frac{k\lambda}{k'} \cos (\gamma- \theta) +\frac{k^2}{k'^2}}+ \frac{q_{TF}}{k'}},$}
\end{equation}
where $\lambda = q/k'$ as in the previous case with $k' = k +q_{TF}$. Similarly, for large $q \to \Lambda$ (where $\Lambda \sim 2\pi/a$ is the ultraviolet cutoff for the lattice-momentum), the integrand in Eq.~\ref{appeq.13} reduces to
\begin{align}\label{appeq.14}
\Sigma(\vk) & \propto k' {\bm \sigma}  \cdot \int^{\frac{\Lambda}{k'}} \frac{\lambda d\lambda}{\lambda} d\theta  (\cos \theta, \sin \theta) \frac{k}{k'}\cos(\gamma-\theta) \nonumber \\
& = {\bm \sigma}  \cdot \int^{\frac{\Lambda}{k}} \frac{d\lambda}{\lambda}  (k\cos \gamma, k\sin \gamma) \nonumber \\
& \xRightarrow[\Lambda \to \infty]{} {\bm \sigma} \cdot \vk  \log \left( \frac{\Lambda}{k+ q_{TF}}\right).
\end{align}
Hence, we obtain the self-energy correction (see Eq.~\ref{eq.8}) as $\Sigma(\vk) = \frac{e^2}{\epsilon_0} \log \left( \frac{\Lambda}{k+q_{TF}}\right) {\bm \sigma} \cdot \vk$. Clearly, the finite range of the screened Coulomb interaction (signaled by non-zero $q_{TF}$) leads to the disappearance of the divergence as in Eq.~\ref{eq.3}.

A similar analysis for the Yukawa potential, would then generate an analogous cutoff in the renormalization as $\Sigma(\vk) = \frac{e^2}{\epsilon_0} \log \left( \frac{\Lambda}{\sqrt{k^2+q^2_{TF}}}\right) {\bm \sigma} \cdot \vk$.

%===========================================================================
\section{Fermionic second-order perturbation theory: Hubbard interaction \label{sec:Apphubbard}}
%===========================================================================

\begin{figure}[t]
\centering
\includegraphics[width=1\linewidth]{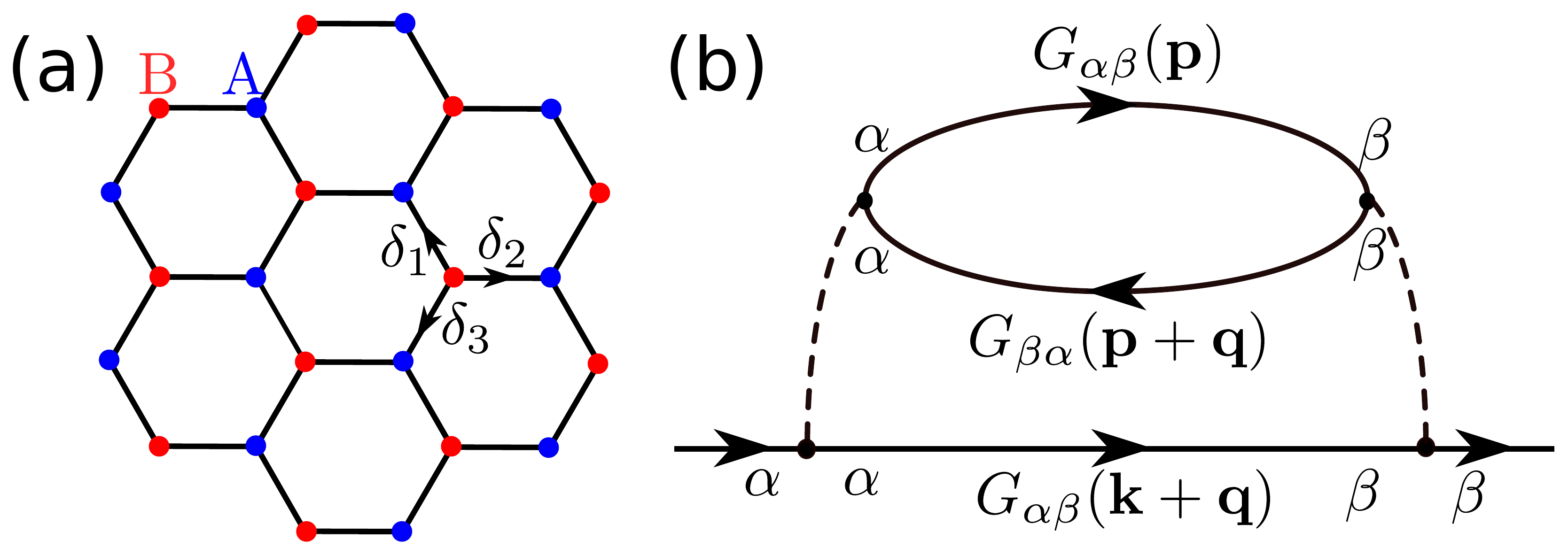}
\caption{(a) The honeycomb lattice for the Hubbard model. Red (blue) symbols signify the inequivalent lattice sites A (B), with $\delta_i$'s being its three nearest-neighbour vectors. (b) The second-order Feynman diagram for the local Hubbard interaction, with $\alpha (\beta)$ representing the sub-lattice indices.}
\label{fig:App_Fig2}
\end{figure}
%===========================================================================

In this part, we outline the detailed analysis of the second-order perturbation theory for the half-filled Hubbard model on a honeycomb lattice. The tight-binding Hamiltonian of Eq.~\ref{eq.10} can be written in momentum-space as
\begin{equation}\label{appeq.15}
H_0 = -t \sum_{\vk} \left(\gamma_\vk a^{\dagger}_{\vk}b_\vk + \gamma^*_\vk b^{\dagger}_\vk a_\vk \right),
\end{equation}
where $a^{\dagger}_\vk (b^{\dagger}_\vk)$ represents the fermionic creations operator on sublattice A(B), with $\gamma_\vk = \sum_{{\bm \delta}_i}e^{i \vk \cdot {\bm \delta}_i}$. Here $\bm \delta_i$ are the three nearest-neighbour vectors on the lattice as shown in Fig.~\ref{fig:App_Fig2}a. Rewriting $\gamma_\vk = |\gamma_\vk|e^{i\phi_\vk}$, we obtain the Green's function in sublattice basis as
\begin{align}\label{appeq.16}
G_{11}(& \vk, i\omega_n) = G_{22}(\vk, i \omega_n)  = \frac{1}{2} \Bigg[ \frac{1}{i \omega_n + \varepsilon_\vk} + \frac{1}{i\omega_n - \varepsilon_\vk} \Bigg] \nonumber \\
& G_{12}(\vk, i \omega_n) = e^{-i\frac{\phi_\vk}{2}} \Bigg[ \frac{\cos \frac{\phi_\vk}{2}}{i \omega_n + \varepsilon_\vk} + \frac{i\sin \frac{\phi_\vk}{2}}{i\omega_n - \varepsilon_\vk} \Bigg]
\end{align}
where $G_{\alpha \beta}$ ($\alpha,\beta = 1,2$ label the two sublattices A, B) corresponds to the Green's function between sublattices $\alpha$ and $\beta$ with the dispersion written as $\varepsilon_\vk = |\gamma_\vk|$ (in unit of the tight-binding parameter $t$). $G_{21}(\vk, i \omega_n) = G_{12}(\vk, i \omega_n \rightarrow -i\omega_n)^{\ast}$. The self-energy (upto second order in $U$) in sublattice basis then follows from the Hubbard interaction as
\begin{align}\label{appeq.17}
& \Sigma_{\uparrow, \alpha \beta} (\vk, i\omega_n)  = \delta_{\alpha \beta} \frac{U T}{N} \sum_{k'} G_{\downarrow, \alpha \alpha} (k') - \frac{UT^2}{N^2} \sum_{k',k''} \nonumber \\
&  G_{\uparrow, \alpha \nu'} (k') G_{\uparrow, \alpha \nu'} (k')G_{\downarrow \alpha \nu''} (k'')  \cdot G_{\downarrow, \nu \alpha} (k''') V^{\nu' \nu'' \beta \nu}_{k',k'',k,k'''},
\end{align}
where the momentum conservation at the vertices implies $k''' = k' +k'' -k$. Here, we used the short-hand notation as $k = (\vk, i\omega_n)$. The Hubbard interaction in Eq.~\ref{appeq.17} has been rewritten as
\begin{equation}\label{appeq.18}
V^{\nu' \nu'' \beta \nu}_{k',k'',k,k'''} = U \delta_{\nu'\nu''}\delta_{\nu' \beta} \delta_{\beta \nu}.
\end{equation}
Using Eq.~\ref{appeq.17} and Eq.~\ref{appeq.18}, the second order self-energy $\Sigma_{\alpha\beta}$ is obtained (see Fig.~\ref{fig:App_Fig2}b for the corresponding Feynman diagram) as
\begin{align}\label{appeq.19}
\Sigma_{\alpha \beta} & = U^2 T \sum_{q} G_{\alpha \beta }(k+q) \chi_{\alpha \beta}(q) \\
\chi_{\alpha \beta}(q) & = - T \sum_{p} G_{\alpha \beta} (p) G_{\beta \alpha} (p+q),
\end{align}
where the spin-index has been suppressed. The polarization function $\chi_{\alpha \beta}(k)$ corresponds to the bubble shown in Fig.~\ref{fig:App_Fig2}b. Using the sublattice representations of the Green's function in Eq.~\ref{appeq.16}, we obtain the four components of the polarization function $\chi_{\alpha \beta}$ as
\begin{strip}
\begin{align}
\chi_{11}(q) &= \chi_{22}(q) = \frac{1}{4} \sum_{\vp} \Bigg[ \frac{f(-\varepsilon_{\vp})-f(-\varepsilon_{\vp'})}{i\epsilon_m + \varepsilon_{\vp'}-\varepsilon_\vp} +\frac{f(-\varepsilon_{\vp})-f(\varepsilon_{\vp'})}{i\epsilon_m - \varepsilon_{\vp'}-\varepsilon_\vp} +\frac{f(\varepsilon_{\vp})-f(-\varepsilon_{\vp'})}{i\epsilon_m + \varepsilon_{\vp'}+\varepsilon_\vp} + \frac{f(\varepsilon_{\vp})-f(\varepsilon_{\vp'})}{i\epsilon_m + \varepsilon_{\vp'}+\varepsilon_\vp} \Bigg],  \\
\chi_{12}(q) &= \sum_{\vp} \Bigg[ \cos \frac{\phi_\vp}{2} \cos \frac{\phi_{\vp'}}{2}\frac{f(-\varepsilon_{\vp})-f(-\varepsilon_{\vp'})}{i\epsilon_m + \varepsilon_{\vp'}-\varepsilon_\vp} + \sin \frac{\phi_\vp}{2} \sin \frac{\phi_{\vp'}}{2}\frac{f(\varepsilon_{\vp})-f(\varepsilon_{\vp'})}{i\epsilon_m - \varepsilon_{\vp'}+\varepsilon_\vp} - \nonumber \\
& \qquad \qquad \qquad \qquad i\cos \frac{\phi_\vp}{2} \sin \frac{\phi_{\vp'}}{2}\frac{f(-\varepsilon_{\vp})-f(\varepsilon_{\vp'})}{i\epsilon_m - \varepsilon_{\vp'}-\varepsilon_\vp} + i \sin \frac{\phi_\vp}{2} \cos \frac{\phi_{\vp'}}{2} \frac{f(\varepsilon_{\vp})-f(-\varepsilon_{\vp'})}{i\epsilon_m + \varepsilon_{\vp'}+\varepsilon_\vp} \Bigg],  \\
\chi_{21}(q) &=  \sum_{\vp} \Bigg[ \cos \frac{\phi_\vp}{2} \cos \frac{\phi_{\vp'}}{2}\frac{f(-\varepsilon_{\vp})-f(-\varepsilon_{\vp'})}{i\epsilon_m + \varepsilon_{\vp'}-\varepsilon_\vp} + \sin \frac{\phi_\vp}{2} \sin \frac{\phi_{\vp'}}{2}\frac{f(\varepsilon_{\vp})-f(\varepsilon_{\vp'})}{i\epsilon_m - \varepsilon_{\vp'}+\varepsilon_\vp} - \nonumber \\
& \qquad \qquad \qquad \qquad  i\sin \frac{\phi_\vp}{2} \cos \frac{\phi_{\vp'}}{2}\frac{f(\varepsilon_{\vp})-f(-\varepsilon_{\vp'})}{i\epsilon_m + \varepsilon_{\vp'}+\varepsilon_\vp} + i \cos \frac{\phi_\vp}{2} \sin \frac{\phi_{\vp'}}{2} \frac{f(-\varepsilon_{\vp})-f(\varepsilon_{\vp'})}{i\epsilon_m - \varepsilon_{\vp'}-\varepsilon_\vp} \Bigg].
\end{align}
\end{strip}
Now, we compute the self-energy, defined in Eq.~\ref{appeq.19}, in the sublattice basis with the $2\times 2$- matrix-valued polarization function. After performing the summation over the relevant Matsubara frequencies we obtain the second order self-energy $\Sigma_{\alpha \beta}(\vk, i\omega_n)$ in the sublattice basis as
\begin{strip}
\begin{align}\label{appeq.21}
\Sigma_{11}(k) & = \Sigma_{22}(k) = \frac{U^2}{8}\sum_{\vp,\vq} \Bigg[ \frac{f(\varepsilon_{\vp'})f(-\varepsilon_{\vp})f(-\varepsilon_{\vk'}) + f(-\varepsilon_{\vp'})f(\varepsilon_{\vp})f(\varepsilon_{\vk'})}{i\omega_n + \varepsilon_{\vk'}-\varepsilon_{\vp'}+\varepsilon_{\vp}} +   \nonumber \\
& \frac{f(-\varepsilon_{\vp'})f(-\varepsilon_{\vp})f(-\varepsilon_{\vk'}) + f(\varepsilon_{\vp'})f(\varepsilon_{\vp})f(\varepsilon_{\vk'})}{i\omega_n + \varepsilon_{\vk'}+\varepsilon_{\vp'}+\varepsilon_{\vp}} + \nonumber \\
&  \frac{f(\varepsilon_{\vp'})f(\varepsilon_{\vp})f(-\varepsilon_{\vk'}) + f(-\varepsilon_{\vp'})f(-\varepsilon_{\vp})f(\varepsilon_{\vk'})}{i\omega_n + \varepsilon_{\vk'}-\varepsilon_{\vp'}-\varepsilon_{\vp}} + \frac{f(-\varepsilon_{\vp'})f(\varepsilon_{\vp})f(-\varepsilon_{\vk'}) + f(\varepsilon_{\vp'})f(-\varepsilon_{\vp})f(\varepsilon_{\vk'})}{i\omega_n + \varepsilon_{\vk'}+\varepsilon_{\vp'}-\varepsilon_{\vp}} + \nonumber \\ &\frac{f(\varepsilon_{\vp'})f(\varepsilon_{\vp})f(\varepsilon_{\vk'}) + f(-\varepsilon_{\vp'})f(\varepsilon_{\vp})f(-\varepsilon_{\vk'})}{i\omega_n - \varepsilon_{\vk'}-\varepsilon_{\vp'}+\varepsilon_{\vp}} + \frac{f(-\varepsilon_{\vp'})f(-\varepsilon_{\vp})f(\varepsilon_{\vk'}) + f(\varepsilon_{\vp'})f(\varepsilon_{\vp})f(-\varepsilon_{\vk'})}{i\omega_n - \varepsilon_{\vk'}+\varepsilon_{\vp'}+\varepsilon_{\vp}} + \nonumber \\
& \frac{f(\varepsilon_{\vp'})f(\varepsilon_{\vp})f(\varepsilon_{\vk'}) + f(-\varepsilon_{\vp'})f(-\varepsilon_{\vp})f(-\varepsilon_{\vk'})}{i\omega_n - \varepsilon_{\vk'}-\varepsilon_{\vp'}-\varepsilon_{\vp}} + \frac{f(-\varepsilon_{\vp'})f(\varepsilon_{\vp})f(\varepsilon_{\vk'}) + f(\varepsilon_{\vp'})f(-\varepsilon_{\vp})f(-\varepsilon_{\vk'})}{i\omega_n - \varepsilon_{\vk'}+\varepsilon_{\vp'}-\varepsilon_{\vp}} \Bigg],
\end{align}
\end{strip}
where $\Sigma_{11}(k)$ and $\Sigma_{22}(k)$ are the identical diagonal entries of the $2\times2$-self-energy  matrix. The off-diagonal entries are even longer and are provided as follows

\begin{strip}
\begin{align}
\Sigma_{12}(k) &=   U^2 \sum_{\vp, \vq} e^{\frac{i}{2} (\phi_{\vp'} -\phi_\vp -\phi_{\vk'})} \Bigg[\cos\frac{\phi_{\vk'}}{2} \cos\frac{\phi_{\vp}}{2} \cos\frac{\phi_{\vp'}}{2}  \frac{f(\varepsilon_{\vp'})f(-\varepsilon_{\vp})f(-\varepsilon_{\vk'}) + f(-\varepsilon_{\vp'})f(\varepsilon_{\vp})f(\varepsilon_{\vk'})}{i\omega_n + \varepsilon_{\vk'}-\varepsilon_{\vp'}+\varepsilon_{\vp}} + \nonumber \\
& \quad \cos\frac{\phi_{\vk'}}{2} \sin\frac{\phi_{\vp}}{2} \sin\frac{\phi_{\vp'}}{2} \frac{f(-\varepsilon_{\vp'})f(\varepsilon_{\vp})f(-\varepsilon_{\vk'}) + f(\varepsilon_{\vp'})f(-\varepsilon_{\vp})f(\varepsilon_{\vk'})}{i\omega_n + \varepsilon_{\vk'}+\varepsilon_{\vp'}-\varepsilon_{\vp}} + \sin\frac{\phi_{\vk'}}{2} \cos\frac{\phi_{\vp}}{2} \sin\frac{\phi_{\vp'}}{2} \times \nonumber \\
& \quad \frac{f(-\varepsilon_{\vp'})f(-\varepsilon_{\vp})f(\varepsilon_{\vk'}) + f(\varepsilon_{\vp'})f(\varepsilon_{\vp})f(-\varepsilon_{\vk'})}{i\omega_n - \varepsilon_{\vk'}+\varepsilon_{\vp'}+\varepsilon_{\vp}} - \sin\frac{\phi_{\vk'}}{2} \sin\frac{\phi_{\vp}}{2} \cos\frac{\phi_{\vp'}}{2} \times \nonumber \\
& \quad \frac{f(\varepsilon_{\vp'})f(\varepsilon_{\vp})f(\varepsilon_{\vk'}) + f(-\varepsilon_{\vp'})f(-\varepsilon_{\vp})f(-\varepsilon_{\vk'})}{i\omega_n - \varepsilon_{\vk'}-\varepsilon_{\vp'}-\varepsilon_{\vp}} - i\cos\frac{\phi_{\vk'}}{2} \cos\frac{\phi_{\vp}}{2} \sin\frac{\phi_{\vp'}}{2} \times \nonumber \\
& \quad \frac{f(-\varepsilon_{\vp'})f(-\varepsilon_{\vp})f(-\varepsilon_{\vk'}) + f(\varepsilon_{\vp'})f(\varepsilon_{\vp})f(\varepsilon_{\vk'})}{i\omega_n + \varepsilon_{\vk'}+\varepsilon_{\vp'}+\varepsilon_{\vp}} + i\cos\frac{\phi_{\vk'}}{2} \sin\frac{\phi_{\vp}}{2} \cos\frac{\phi_{\vp'}}{2} \times \nonumber \\
& \quad \frac{f(\varepsilon_{\vp'})f(\varepsilon_{\vp})f(-\varepsilon_{\vk'}) + f(-\varepsilon_{\vp'})f(-\varepsilon_{\vp})f(\varepsilon_{\vk'})}{i\omega_n + \varepsilon_{\vk'}-\varepsilon_{\vp'}-\varepsilon_{\vp}} + i\sin\frac{\phi_{\vk'}}{2} \cos\frac{\phi_{\vp}}{2} \cos\frac{\phi_{\vp'}}{2} \times \nonumber \\
& \quad \frac{f(\varepsilon_{\vp'})f(-\varepsilon_{\vp})f(\varepsilon_{\vk'}) + f(-\varepsilon_{\vp'})f(\varepsilon_{\vp})f(-\varepsilon_{\vk'})}{i\omega_n - \varepsilon_{\vk'}-\varepsilon_{\vp'}+\varepsilon_{\vp}} + \nonumber \\
&  \quad i\sin\frac{\phi_{\vk'}}{2} \sin\frac{\phi_{\vp}}{2} \sin\frac{\phi_{\vp'}}{2} \frac{f(-\varepsilon_{\vp'})f(\varepsilon_{\vp})f(\varepsilon_{\vk'}) + f(\varepsilon_{\vp'})f(-\varepsilon_{\vp})f(-\varepsilon_{\vk'})}{i\omega_n - \varepsilon_{\vk'}+\varepsilon_{\vp'}-\varepsilon_{\vp}} \Bigg], \\
\Sigma_{21}(k) &=   U^2 \sum_{\vp, \vq} e^{\frac{i}{2} (\phi_{\vp} +\phi_{\vk'} -\phi_{\vp'})} \Bigg[\cos\frac{\phi_{\vk'}}{2} \cos\frac{\phi_{\vp}}{2} \cos\frac{\phi_{\vp'}}{2}  \frac{f(\varepsilon_{\vp'})f(-\varepsilon_{\vp})f(-\varepsilon_{\vk'}) + f(-\varepsilon_{\vp'})f(\varepsilon_{\vp})f(\varepsilon_{\vk'})}{i\omega_n + \varepsilon_{\vk'}-\varepsilon_{\vp'}+\varepsilon_{\vp}} + \nonumber \\
& \quad \cos\frac{\phi_{\vk'}}{2} \sin\frac{\phi_{\vp}}{2} \sin\frac{\phi_{\vp'}}{2} \frac{f(-\varepsilon_{\vp'})f(\varepsilon_{\vp})f(-\varepsilon_{\vk'}) + f(\varepsilon_{\vp'})f(-\varepsilon_{\vp})f(\varepsilon_{\vk'})}{i\omega_n + \varepsilon_{\vk'}+\varepsilon_{\vp'}-\varepsilon_{\vp}} - \sin\frac{\phi_{\vk'}}{2} \sin\frac{\phi_{\vp}}{2} \cos\frac{\phi_{\vp'}}{2} \times \nonumber \\
& \quad \frac{f(\varepsilon_{\vp'})f(\varepsilon_{\vp})f(\varepsilon_{\vk'}) + f(-\varepsilon_{\vp'})f(-\varepsilon_{\vp})f(-\varepsilon_{\vk'})}{i\omega_n - \varepsilon_{\vk'}-\varepsilon_{\vp'}-\varepsilon_{\vp}} + \sin\frac{\phi_{\vk'}}{2} \cos\frac{\phi_{\vp}}{2} \sin\frac{\phi_{\vp'}}{2} \times \nonumber \\
& \quad \frac{f(-\varepsilon_{\vp'})f(-\varepsilon_{\vp})f(\varepsilon_{\vk'}) + f(\varepsilon_{\vp'})f(\varepsilon_{\vp})f(-\varepsilon_{\vk'})}{i\omega_n - \varepsilon_{\vk'}+\varepsilon_{\vp'}+\varepsilon_{\vp}} - i\cos\frac{\phi_{\vk'}}{2} \sin\frac{\phi_{\vp}}{2} \cos\frac{\phi_{\vp'}}{2} \times \nonumber \\
& \quad \frac{f(\varepsilon_{\vp'})f(\varepsilon_{\vp})f(-\varepsilon_{\vk'}) + f(-\varepsilon_{\vp'})f(-\varepsilon_{\vp})f(\varepsilon_{\vk'})}{i\omega_n + \varepsilon_{\vk'}-\varepsilon_{\vp'}-\varepsilon_{\vp}} + i\cos\frac{\phi_{\vk'}}{2} \cos\frac{\phi_{\vp}}{2} \sin\frac{\phi_{\vp'}}{2} \times \nonumber \\
& \quad \frac{f(-\varepsilon_{\vp'})f(-\varepsilon_{\vp})f(-\varepsilon_{\vk'}) + f(\varepsilon_{\vp'})f(\varepsilon_{\vp})f(\varepsilon_{\vk'})}{i\omega_n + \varepsilon_{\vk'}+\varepsilon_{\vp'}+\varepsilon_{\vp}} - i\sin\frac{\phi_{\vk'}}{2} \cos\frac{\phi_{\vp}}{2} \cos\frac{\phi_{\vp'}}{2} \times \nonumber \\
& \quad \frac{f(\varepsilon_{\vp'})f(-\varepsilon_{\vp})f(\varepsilon_{\vk'}) + f(-\varepsilon_{\vp'})f(\varepsilon_{\vp})f(-\varepsilon_{\vk'})}{i\omega_n - \varepsilon_{\vk'}-\varepsilon_{\vp'}+\varepsilon_{\vp}} - \nonumber \\
& \quad i\sin\frac{\phi_{\vk'}}{2} \sin\frac{\phi_{\vp}}{2} \sin\frac{\phi_{\vp'}}{2} \frac{f(-\varepsilon_{\vp'})f(\varepsilon_{\vp})f(\varepsilon_{\vk'}) + f(\varepsilon_{\vp'})f(-\varepsilon_{\vp})f(-\varepsilon_{\vk'})}{i\omega_n - \varepsilon_{\vk'}+\varepsilon_{\vp'}-\varepsilon_{\vp}} \Bigg].
\end{align}
\end{strip}
We have utilized the following identity where $f(x) (b(x))$ corresponds to the Fermi-Dirac (Bose-Einstein) distributions,
\begin{strip}
\begin{equation}\label{appeq.20}
\Big[f(y)-f(x)\Big] \Big[b(x-y) + f(z)\Big]  = f(-x)f(y)f(z) + f(x)f(-y)f(-z).
\end{equation}
\end{strip}
These long expressions for the self-energy in Eqs.~\ref{appeq.21} simplify at zero temperature. We therefore obtain the diagonal entries as
\begin{strip}
\begin{align}\label{appeq.22}
\Sigma(\vk,\omega) & = \frac{U^2}{8} \sum_{\vp, \vq} \Bigg[\frac{1}{\omega -\varepsilon_{\vk'} - \varepsilon_{\vp'} -\varepsilon_{\vp} + i \delta} + \frac{1}{\omega +\varepsilon_{\vk'} + \varepsilon_{\vp'} +\varepsilon_{\vp} + i \delta}\Bigg] 
\end{align}
\end{strip}
where we relabeled the identical diagonal entries $\Sigma_{11}(k)$ and $\Sigma_{22}(k)$ as $\Sigma(k)$ and performed the analytical continuation to real frequency as $i\omega_n \to \omega +i \delta$. The off-diagonal entries are similarly simplied as
\begin{strip}
\begin{align}
\Sigma_{12}(\vk,\omega) & =- U^2 \sum_{\vp, \vq} \Bigg[ \sin \frac{\phi_{\vk'}}{2} \sin \frac{\phi_{\vp}}{2} \cos \frac{\phi_{\vp'}}{2} \frac{e^{\frac{i}{2}(\phi_{\vp'} -\phi_\vp -\phi_{\vk'})}}{\omega -\varepsilon_{\vk'}-\varepsilon_{\vp'} -\varepsilon_\vp +i \delta} + \nonumber \\
& \qquad \qquad \quad i\cos \frac{\phi_{\vk'}}{2} \cos \frac{\phi_{\vp}}{2} \sin \frac{\phi_{\vp'}}{2} \frac{e^{\frac{i}{2}(\phi_{\vp'} -\phi_\vp -\phi_{\vk'})}}{\omega +\varepsilon_{\vk'}+\varepsilon_{\vp'} +\varepsilon_\vp + i \delta}\Bigg] \nonumber \\
\Sigma_{21}(\vk,\omega) & =- U^2 \sum_{\vp, \vq} \Bigg[ \sin \frac{\phi_{\vk'}}{2} \sin \frac{\phi_{\vp}}{2} \cos \frac{\phi_{\vp'}}{2} \frac{e^{\frac{i}{2}(\phi_{\vp} +\phi_{\vk'} -\phi_{\vp'})}}{\omega -\varepsilon_{\vk'}-\varepsilon_{\vp'} -\varepsilon_\vp + i \delta} - \nonumber \\
& \qquad \qquad \quad i\cos \frac{\phi_{\vk'}}{2} \cos \frac{\phi_{\vp}}{2} \sin \frac{\phi_{\vp'}}{2} \frac{e^{\frac{i}{2}(\phi_{\vp} +\phi_{\vk'} -\phi_{\vp'})}}{\omega +\varepsilon_{\vk'}+\varepsilon_{\vp'} +\varepsilon_\vp + i \delta}\Bigg],
\end{align}
\end{strip}

Now, we focus on the spectral function for this effective two-band system as
\begin{equation}\label{appeq.23}
A(\vk,\omega) = - \text{Im} \Tr \big[ G(\vk,\omega)\big],
\end{equation}
where the matrix Green's function is defined as
\begin{equation}\label{appeq.24}
G^{-1}(\vk,\omega) = \begin{pmatrix}
\omega  - \Sigma(\vk, \omega) & \gamma_\vk -\Sigma_{12}(\vk, \omega) \\
\gamma^*_\vk - \Sigma_{21}(\vk, \omega) & \omega - \Sigma(\vk, \omega)
\end{pmatrix}.
\end{equation}

%===========================================================================
\section{Bosonic second-order perturbation theory: Coulomb interaction \label{sec:Appcoulomb}}
%===========================================================================

\begin{figure}[t]
\centering
\includegraphics[width=0.8\linewidth]{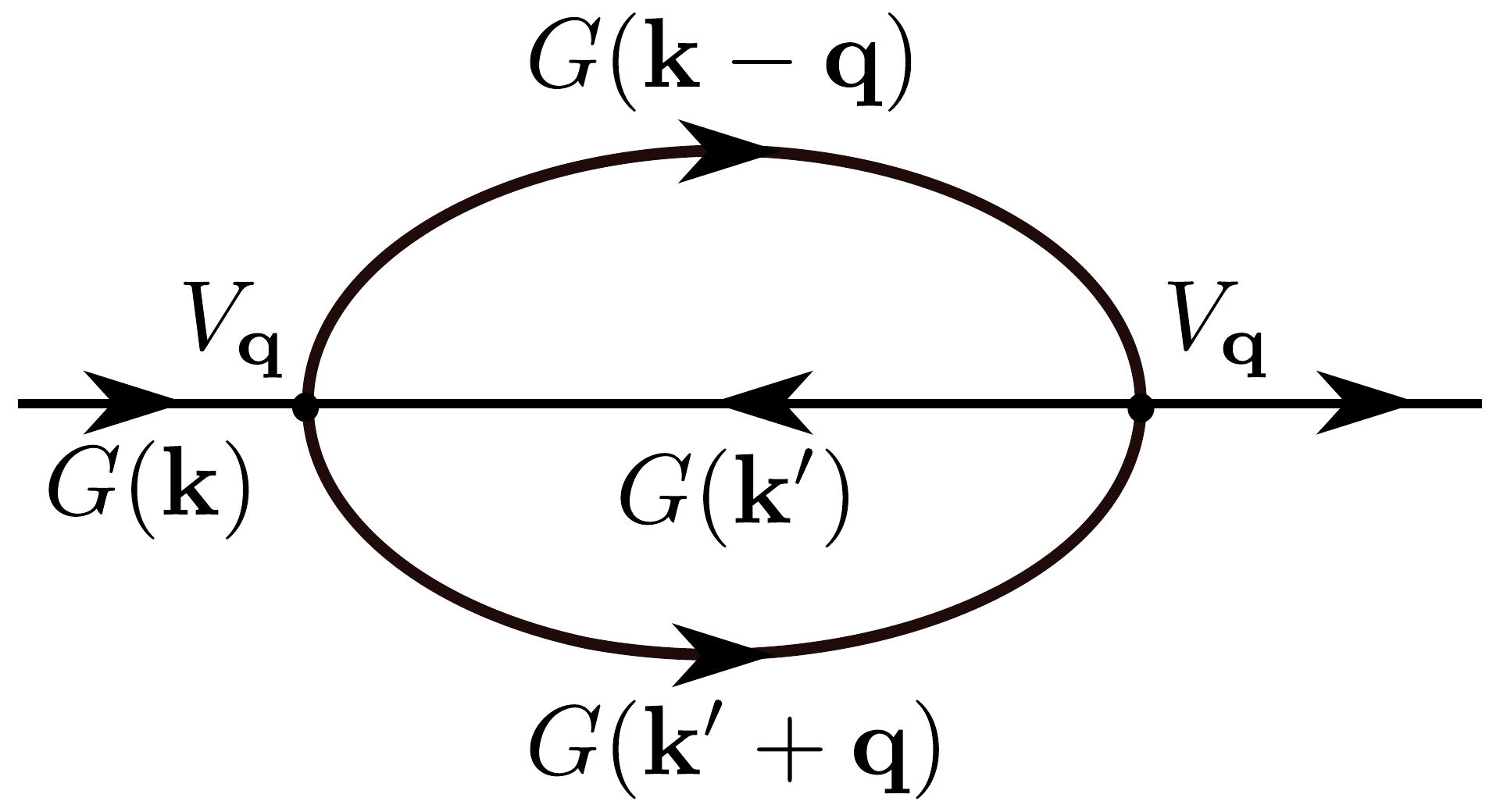}
\caption{The self-energy diagram in the second-order perturbation theory for the Dirac bosons interacting via the long-range Coulomb interaction.}
\label{fig:App_Fig4}
\end{figure}

 In this part, we examine the Hamiltonian (see Eq.~\ref{eq.24}) for the Dirac bosons interacting via long-range Coulomb interaction. Rewriting the bosonic Hamiltonian in the diagonal basis we obtain
\begin{align}\label{app.25}
H & = \sum_\vk \Big[ \varepsilon^u_\vk u^{\dagger}_\vk u_\vk + \varepsilon^d_\vk d^{\dagger}_\vk d_\vk \Big] + \mathcal{H}_{\text{int}}, \\
\mathcal{H}_{\text{int}} & = \sum_{\vq,\vk,\vk'} V_\vq \big[ a^{\dagger}_{\vk-\vq}a^{\dagger}_{\vk'+\vq}a_{\vk'}a_\vk + b^{\dagger}_{\vk-\vq}b^{\dagger}_{\vk'+\vq}b_{\vk'}b_\vk  \nonumber \\
& \qquad \qquad \qquad \qquad \qquad + a^{\dagger}_{\vk-\vq}b^{\dagger}_{\vk'+\vq}b_{\vk'}a_\vk  \Big],
\end{align}
where $a^{\dagger}_\vk$($b^{\dagger}_\vk$) is the creation operator for bosons on sublattice A(B) and $u^{\dagger}_\vk$($d^{\dagger}_\vk$) corresponds to bosonic operators in the diagonal basis with the dispersion $\varepsilon^{(u/d)}_\vk = t(3 \pm |\gamma_\vk|)$. Here, $\gamma_\vk$ is identical to the one discussed in the previous Appendix and $V_\vq = \frac{{e^*}^2}{\epsilon_0} \frac{1}{|\vq|}$. The diagonal and sub-lattice operators are related to each other via
\begin{align}\label{appeq.26}
a_\vk & = e^{i\frac{\phi_\vk}{2}} \frac{\left( u_\vk + d_\vk \right)}{\sqrt{2}},\;  b_\vk  = e^{-i\frac{\phi_\vk}{2}} \frac{\left( -u_\vk + d_\vk \right)}{\sqrt{2}},
\end{align}
where $\gamma_\vk = |\gamma_\vk| e^{i\phi_\vk}$. Next, we rewrite the interacting Hamiltonian in the diagonal basis as
\begin{strip}
\begin{align}\label{appeq.27}
\mathcal{H}_{\text{int}} & =  \sum_{\vq,\vk,\vk'} V_\vq \Bigg[ A(\Phi,\Psi) u^{\dagger}_{\vk-\vq}u^{\dagger}_{\vk'+\vq}u_{\vk'}u_\vk + i\sin\Phi u^{\dagger}_{\vk-\vq}u^{\dagger}_{\vk'+\vq}u_{\vk'}d_\vk + A(\Phi,\Psi) u^{\dagger}_{\vk-\vq}u^{\dagger}_{\vk'+\vq}d_{\vk'}d_\vk\nonumber \\
&  + i\sin\Phi u^{\dagger}_{\vk-\vq}d^{\dagger}_{\vk'+\vq}u_{\vk'}u_\vk + 2\cos\Phi u^{\dagger}_{\vk-\vq}d^{\dagger}_{\vk'+\vq}u_{\vk'}d_\vk + i\sin\Phi u^{\dagger}_{\vk-\vq}d^{\dagger}_{\vk'+\vq}d_{\vk'}d_\vk + A(\Phi,\Psi) d^{\dagger}_{\vk-\vq}d^{\dagger}_{\vk'+\vq}d_{\vk'}d_\vk + h.c. \Bigg]  \nonumber \\
\end{align}
\end{strip}
where the two angles are defined as $\Phi = (\phi_\vk + \phi_{\vk'} - \phi_{\vk-\vq} -\phi_{\vk'+\vq})/2$ and $\Psi = (\phi_{\vk'+\vq} + \phi_{\vk} - \phi_{\vk-\vq} -\phi_{\vk'})/2$ and $A(\Phi,\Psi)  = \left(\frac{1}{2}\cos\Phi-\frac{1}{4}\cos\Psi\right)$. However, not all of these scattering channels are relevant in the low temperature window due to the kinematic consideration. We compute the correction to the bare Dirac spectrum within a second-order perturbation theory in the strength of the Coulomb interaction following the Feynman diagram (see Fig.\ref{fig:App_Fig4}) and obtain the former as
\begin{equation}\label{appeq.28}
\Sigma(\omega,\vk) = \sum_{\vk',\vq}\frac{|V_{\vq,\vk,\vk'}|^2 F(\vk;\vk',\vq)}{\omega +\varepsilon_{\vk'} - \varepsilon_{\vk-\vq}-\varepsilon_{\vk'+\vq}+i\delta};
\end{equation}
where $V_{\vq,\vk,\vk'}$ is the relevant scattering strength for the particular channel following Eq.\ref{appeq.27} and $F(\vk;\vk',\vq)$ is the corresponding thermodynamic factor. The later is obtained after performing the summation over the Matsubara frequency as~\cite{Saikat}
\begin{align}\label{appeq.29}
F(\vk;\vk',\vq) = & \Big[1+b(\varepsilon_{\vk-\vq}) \Big] \Big[ 1 +b(\varepsilon_{\vk'+\vq}) \Big] b(\varepsilon_{\vk'})- \nonumber \\
& b(\varepsilon_{\vk-\vq})b(\varepsilon_{\vk'+\vq}) \Big[1+b(\varepsilon_{\vk'}) \Big],
\end{align}
where $b(x)$ corresponds to Bose-Einstein distribution function. On a closer inspection, we observe that the first term in the above expression corresponds to the direct scattering process, i.e. where the original boson with momentum $\vk$ and the thermal boson with momentum $\vk'$ scatter into bosons with momenta $\vk-\vq$ and $\vk'+\vq$, whereas the second term to the reverse scattering process. For very low temperature $T \ll t$, the direct scattering process dominates over the reverse one, and $F(\vk;\vk',\vq)$ can be approximated as $F(\vk;\vk',\vq) \approx b(\varepsilon^d_{\vk'})$. Following this approximation, the on-
%===========================================================================
%===========================================================================
shell self-energy further simplifies as
\begin{equation}\label{appeq.30}
\Sigma(\vk) \cong \sum_{\vk',\vq}\frac{|V_{\vq,\vk,\vk'}|^2 b(\varepsilon^d_{\vk'})}{\varepsilon_\vk - \varepsilon_{\vk-\vq}-\varepsilon_{\vq}+i\delta}.
\end{equation}
At the same time, only three scattering processes among the eight different scattering channels (see Eq.\ref{appeq.27}) are relevant at low temperature due to kinematic considerations. These processes are labeled as $u \rightarrow u+d$, $u \rightarrow d+d$ and $d \rightarrow d+ d$, where an ``up'' (``down'') boson effectively scatters with a ``\textit{ghost-down}'' boson (\textit{thermally excited bosons around the bottom of the bosonic band-structure as illustrated in Fig.\ref{fig:Spectrum}a}) into an ``up'' (``down'') and ``down'' bosons.

%===========================================================================
\section{Interacting Dirac bosons: Analysis \label{sec:Correction}}
%===========================================================================

\begin{figure}[b]
\centering
\includegraphics[width=1\linewidth]{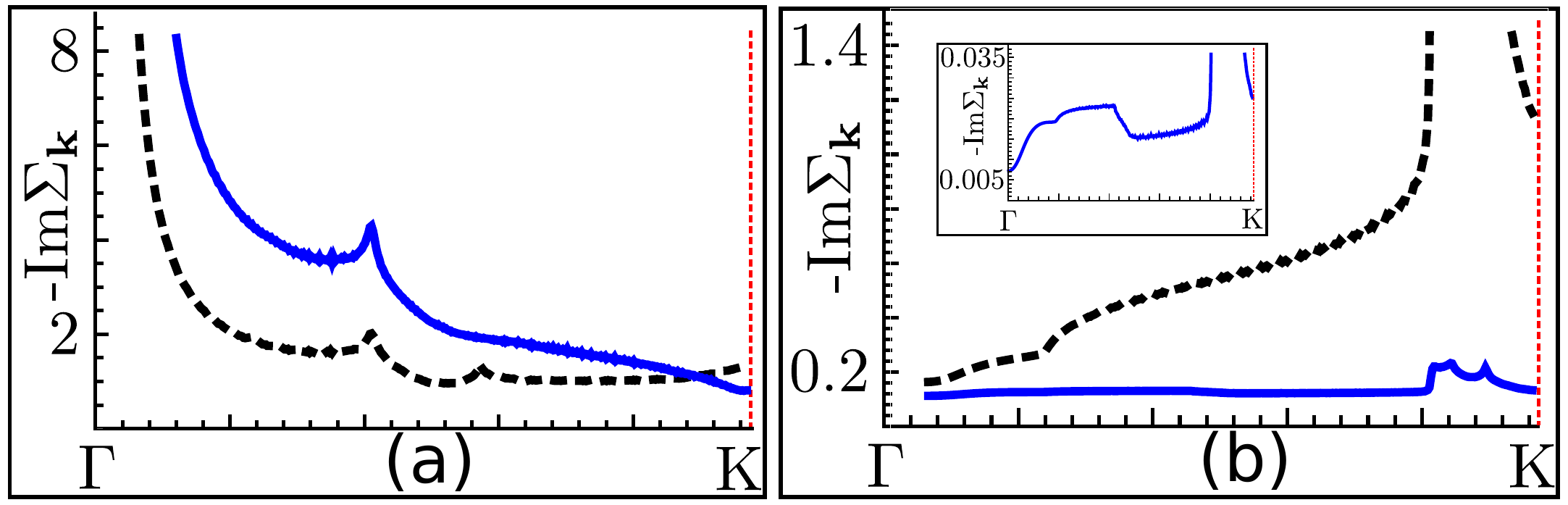}
\caption{The scattering rates (from Eq.~\ref{eq.28} with the screened Coulomb interaction) for the ``up'' (a) and ``down'' (b) bosons in the unit of ${e^*}^4/\epsilon^2_0 t^2$ as a function of the momenta along the $\Gamma$ $\rightarrow$ K point in the Brillouin zone at temperature $T = 0.1 t$. \textit{Dashed lines }\--- without the phase-dependent matrix elements; \textit{Solid lines} \--- with the phase-dependent matrix elements. \textit{Inset} \--- Magnified illustration of the solid line in panel (b).}
\label{fig:App_Fig5}
\end{figure}

In this appendix, we analyze the momentum-dependence of the self-energy correction (see Eq.~\ref{eq.28}) in detail. First of all, the bare Coulomb interaction for a charged particle always is screened. Hence, we incorporate a similar modification (see Sec.~\ref{sc_coulomb_fermi}) to the potential $V_{\vq}$ as $V^{\text{S}}_{\vq} = \frac{{e^*}^2}{\epsilon_0}\frac{1}{q+q_{TF}}$. Of course, the phase factors that emerge as a result of the change of basis from the sublattice to the diagonal basis as in Eq.~\ref{appeq.27}, remain the same. The scattering rates for the relevant three channels are shown in Fig.~\ref{fig:App_Fig5}a,b (solid lines). Note that there are two channels for the scattering of ``up'' bosons compared to only one channel for ``down'' bosons. The scattering rates for the corresponding two channels of the ``up'' bosons are added and illustrated in Fig.~\ref{fig:App_Fig5}a.

%===========================================================================
%===========================================================================

However, the same prescription for the self-energy (Eq.~\ref{eq.28}) leads to a negative correction for the down-band. This feature dictates that the renormalized energy for the Goldstone modes (down bosons near the $\Gamma$ point) becomes negative, which is unphysical. At this point, we notice that our starting bosonic tight-binding model (see Eq.~\ref{eq.24}) corresponds to the collective excitations of some underlying particles such as spins that give rise to a similar magnon model. Hence, the matrix elements in Eq.~\ref{appeq.27} are an effective description and are not the full theory. We believe that the emergence of this negative bosonic self-energy is an artifact of such an incomplete model.

 To cure such an unphysical result, we focus on understanding the effect of the phase-dependent matrix elements. The relevant scattering rates for the up and down band (both with and without the corresponding phase factors in the matrix elements in Eq.`\ref{appeq.27} are shown in Fig.~\ref{fig:App_Fig5}. The results are plotted only along the $\Gamma$ $\rightarrow$ K, instead of the full Brillouin zone to illustrate the renormalization around the Dirac point. The solid lines are the results of the exact matrix elements derived in Eq.~\ref{appeq.27}, whereas the dashed lines are the results without the phase factors. Please note that the screened Coulomb interaction is not modified in both the cases, only the phase factors are dropped in the case for Fig.~\ref{fig:App_Fig5}b. Apart from the overall magnitude, the scattering rates are almost identical (\textit{Note: There is an additional kink in Fig.~\ref{fig:App_Fig5}a for the up-band}). This feature dictates that the phase factors do not play that crucial role in the specific momentum-dependence of the scattering rates. Motivated by this fact, we modify the phase-dependent part of all the matrix elements in Eq.~\ref{appeq.27} as
\begin{align}\label{appeq.31}
V_{\vq,\vk}  = \frac{{e^*}^2}{\epsilon_0}\frac{1}{q+q_{\text{TF}}} &\Big[1-\exp \left(\frac{|\vk|}{\lambda}\right) \Big]  \Big[1-\exp \left(\frac{|\vq|}{\lambda}\right) \Big]\cdot \nonumber \\
 &  \Big[1-\exp \left(\frac{|\vk-\vq|}{\lambda} \right) \Big],
\end{align}
where $\lambda$ is a constant. Now, we substitute this in Eq.~\ref{eq.28} to compute the renormalization of the band structure. Without aiming for a numerical estimate, we focus here on the overall trend around the Dirac crossing point.

\begin{figure}[t]
\centering
\includegraphics[width=1\linewidth]{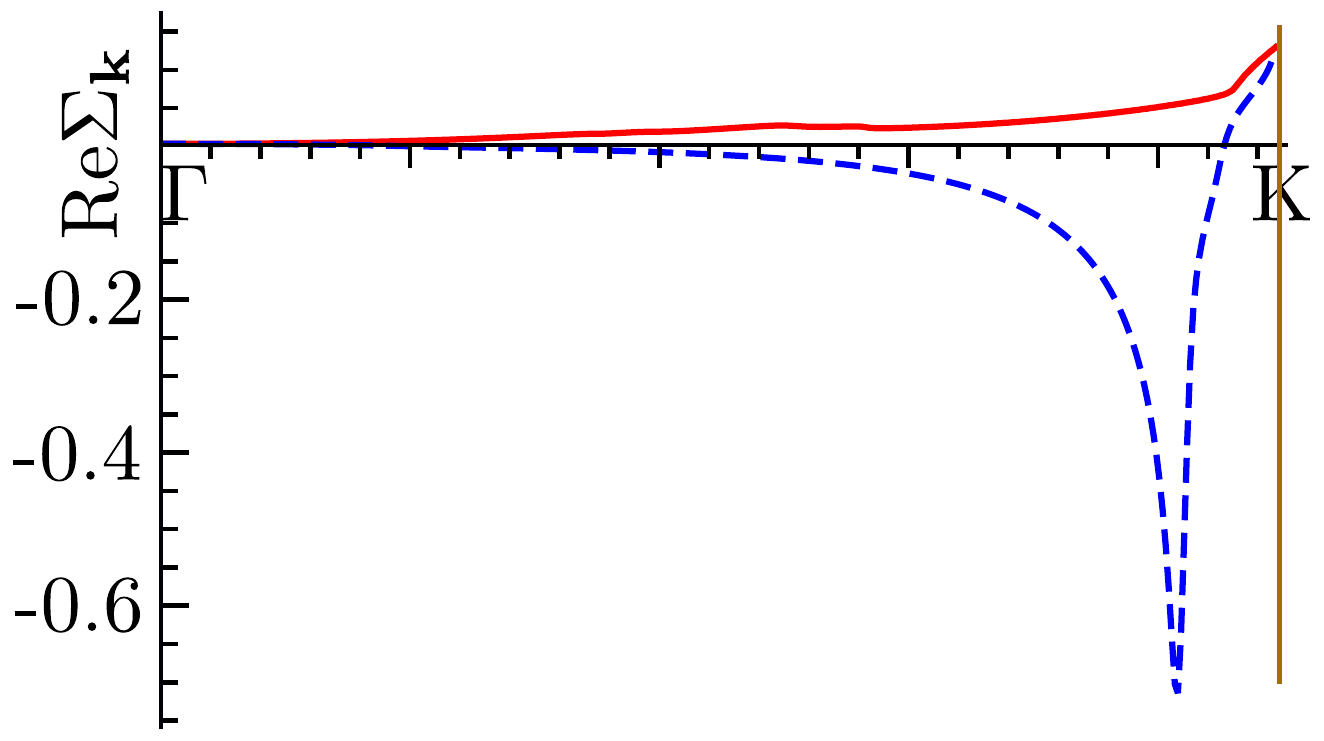}
\caption{The real part of the self-energy (from Eq.~\ref{eq.28} with the screened Coulomb interaction as in Eq.~\ref{appeq.31}) for the ``up'' (solid line) and ``down'' (dashed line) bosons in the unit of ${e^*}^4/\epsilon^2_0 t^2$ as a function of the momenta along the $\Gamma$ $\rightarrow$ K point in the Brillouin zone at temperature $T = 0.1 t$.}
\label{fig:App_Fig6}
\end{figure}

The real part of the self-energy for the two bands are consequently computed and illustrated in Fig.~\ref{fig:App_Fig6}.  Once again we stress that we do not aim for any numerical estimate from our model but only focus on evaluating the overall trend of the energy renormalization around the Dirac crossing point. Fig.~\ref{fig:App_Fig6} clearly dictates that the Dirac velocity increases for the Coulomb interaction. At the same time, we observe that the weak Coulomb interaction does not open any gap at the Dirac point. This result is directly comparable to its fermionic counterpart, where we also observed an increased Dirac velocity and no gap opening.

%===========================================================================
%===========================================================================

%==================================================================================================================
\section*{References}
\bibliographystyle{iopart-num}
\bibliography{References}

%%%%%%%%%%%%%%%%%%%%%%%%%%%%%%%%%%%%%%%%%
\end{document}